\documentclass[usenatbib,useAMS]{mn2e}

\usepackage{times}
\usepackage{amssymb}
\usepackage{amsmath}
\usepackage{graphicx}
\usepackage{bmpsize}
\usepackage{epstopdf}

\newcommand{\be}{\begin{equation}}
\newcommand{\ba}{\begin{eqnarray}}
\newcommand{\ee}{\end{equation}}
\newcommand{\ea}{\end{eqnarray}}

\newcommand{\url}{\tt}%
\def\gtsima{$\; \buildrel > \over \sim \;$}
\def\ltsima{$\; \buildrel < \over \sim \;$}
\def\gsim{\lower.5ex\hbox{\gtsima}}
\def\lsim{\lower.5ex\hbox{\ltsima}}
\def\simgt{\lower.5ex\hbox{\gtsima}}
\def\simlt{\lower.5ex\hbox{\ltsima}}
\def\simpr{\lower.5ex\hbox{\prosima}}

\def\Lya{Ly~$\alpha$}

\def\simless{\mathbin{\lower 3pt\hbox
   {$\rlap{\raise 5pt\hbox{$\char'074$}}\mathchar''7218$}}}  
\def\simgreat{\mathbin{\lower 3pt\hbox
   {$\rlap{\raise 5pt\hbox{$\char'076$}}\mathchar''7218$}}}  

\begin{document}

\title[X-ray Heating during the Cosmic Dawn]{Simulating the Impact of X-ray Heating during the Cosmic Dawn}
\author[H. E. Ross, et al.]
{Hannah E. Ross,$^{1,2}$ \thanks{email: H.Ross@sussex.ac.uk}
Keri L. Dixon$^{1}$, 
Ilian T. Iliev$^{1}$ and
Garrelt Mellema$^{2}$
\\
$^1$ Astronomy Centre, Department of Physics \& Astronomy, Pevensey III Building, University of Sussex, Falmer, Brighton, BN1 9QH, \\United Kingdom
\\
$^2$ Department of Astronomy \& Oskar Klein Centre, AlbaNova, Stockholm University, SE-106 91 Stockholm, Sweden\\
}
\date{\today}
\pubyear{2016} \volume{000} \pagerange{1}
\twocolumn 

\voffset-.6in
\maketitle\label{firstpage}

\begin{abstract}
Upcoming observations of the 21-cm signal from the Epoch of Reionization will soon provide the first direct detection of this era. This signal is influenced by many astrophysical effects, including long range X-ray heating of the intergalactic gas. During the preceding Cosmic Dawn era the impact of this heating on the 21-cm signal is particularly prominent, especially before spin temperature saturation. We present the largest-volume (349\,Mpc comoving=244~$h^{-1}$Mpc) full numerical radiative transfer simulations to date of this epoch that include the effects of helium and multi-frequency heating, both with and without X-ray sources. We show that X-ray sources contribute significantly to early heating of the neutral intergalactic medium and, hence, to the corresponding 21-cm signal. The inclusion of hard, energetic radiation yields an earlier, extended transition from absorption to emission compared to the stellar-only case. The presence of X-ray sources decreases the absolute value of the mean 21-cm differential brightness temperature. These hard sources also significantly increase the 21-cm fluctuations compared the common assumption of temperature saturation. The 21-cm differential brightness temperature power spectrum is initially boosted on large scales, before decreasing on all scales. Compared to the case of the cold, unheated intergalactic medium, the signal has lower rms fluctuations and increased non-Gaussianity, as measured by the skewness and kurtosis of the 21-cm probability distribution functions. Images of the 21-cm signal with resolution around 11~arcmin still show fluctuations well above the expected noise for deep integrations with the SKA1-Low, indicating that direct imaging of the X-ray heating epoch could be feasible.
\end{abstract}

\begin{keywords}
cosmology: theory --- radiative transfer --- reionization ---
intergalactic medium --- large-scale structure of universe ---
galaxies: formation 
\end{keywords}

\section{INTRODUCTION}

The Epoch of Reionization (EoR), a major global phase transition in which the neutral hydrogen in the Universe transitioned from almost neutral to largely ionized, remains one of the cosmological eras least constrained by observations. Although no direct measurements of this transition currently exist, multiple observations indicate reionization is completed by $z \approx 5.7$ and possibly earlier. These observations include high-redshift quasar spectra \citep[e.g.][]{Fan2006,McGreer2015}, the decrease in the fraction of Lyman~$\alpha$ (\Lya) emitting galaxies \citep[e.g.][]{Stark2011,Schenker2012, Pentericci2014, Tilvi2014}, and
measurements of the temperature of the intergalactic medium \citep[IGM; e.g.][]{Theuns2002, Raskutti2012, Bolton2012}. The start
of substantial reionization is constrained by the Thomson optical depth measured from the anisotropies and polarisation of the Cosmic Microwave Background, CMB \citep[e.g.][]{Komatsu2011, Planck2015,   Planck2016}. \cite{Planck2016} find that the Universe was less than 10 per cent ionized at z $\approx$ 10, the average redshift at which reionization would have taken place if it had been an instantaneous process to be in the range $7.8 \leq z \leq 8.8$, and an upper limit for the duration of the process is $\Delta z < 2.8$.

At high redshifts, 21-cm radiation from hydrogen atoms in the I: contains a treasure trove of information about the physical conditions both during the EoR and the preceding epochs. In particular, the 21-cm signal probes the \textit{Dark Ages}, the epoch after recombination during which the formation of baryonic large scale structure began and the \textit{Cosmic Dawn}, the period of preheating from the first ionizing sources before reionization was significantly underway. Several experiments are attempting to measure the 21-cm signal from the EoR using low-frequency radio interferometry. These include the ongoing GMRT\footnote{\url{http://gmrt.ncra.tifr.res.in/}}, LOFAR\footnote{\url{http://www.lofar.org/}}, MWA\footnote{\url{http://www.mwatelescope.org/}}, and PAPER\footnote{\url{http://eor.berkeley.edu/}} and the future HERA\footnote{\url{http://reionization.org/}} and SKA\footnote{\url{https://www.skatelescope.org/}}.

The main sources powering reionization are likely early galaxies, with Population III (Pop.~III; metal-free) and Population II (Pop.~II; metal-enriched) stars providing the bulk of ionizing photons. However, sources of higher energy X-ray photons may also be present, contributing non-trivially to the photon budget. Although their abundance is uncertain, high-mass X-ray binaries (HMXBs) likely exist throughout reionization \citep{Glover2003}. Other hard radiation sources, such as QSOs and supernovae, may have also contributed. Very little is known about these objects in terms of their abundances, clustering, evolution, and spectra, especially at these high redshifts.

The high-energy photons from these hard radiation sources have a much smaller cross section for interaction with atoms and, hence, far longer mean free paths than lower energy ionizing photons. Therefore, these photons are able to penetrate significantly further into the neutral IGM. While not sufficiently numerous to contribute significantly to the ionization of the IGM (although recently there has been some debate about the level of contribution of quasars, \citep[e.g.][]{Khaire2016}), their high energies result in a non-trivial amount of heating. Along with variations in the early \Lya\ background, variations in the temperature of the neutral IGM caused by this non-uniform heating constitute an important source of 21-cm fluctuations before large-scale reionization patchiness develops \citep[see e.g.][for a detailed discussion]{2012RPPh...75h6901P}.

Once a sufficient \Lya\  background due to stellar radiation has been established in the IGM, the spin temperature of neutral hydrogen will be coupled to the kinetic temperature, $T_\mathrm{K}$, due to the Wouthuysen-Field (WF) effect. The 21-cm signal is then expected to appear initially in absorption against the CMB, as the CMB temperature ($T_{\mathrm{CMB}}$) is greater than the spin temperature of the gas. Once the first sources have heated the IGM and brought the spin temperature,$T_\mathrm{s}$, above $T_{\mathrm{CMB}}$, the signal transitions into emission (see Section~\ref{sec:dbtTheory} for more details). The timing and duration of this transition are highly sensitive to the type of sources present, as they determine the quantity and morphology of the heating of the IGM \citep[e.g.][]{Pritchard2007,Baek2010,Mesinger2013,Fialkov2014,2014MNRAS.443..678P,Ahn2015}.

Considerable theoretical work regarding the impact of X-ray radiation on the thermal history of reionization and the future observational signatures exists. Attempts have been made to understand the process analytically \citep[e.g.][]{Glover2003,Furlanetto2004}, semi-numerically \citep[see e.g.][]{Santos2010,Mesinger2013, Fialkov2014, Knevitt2014}, and numerically \citep[eg.][]{Baek2010,Xu2014,Ahn2015}. However, due to the computationally challenging, multi-scale nature of the problem, numerical simulations have not yet been run over a sufficiently large volume -- a few hundred comoving Mpc per side -- to properly account for the patchiness of reionization \citep{Iliev2013}, while at the same time resolving the ionizing sources. 

In this paper, we present the first full numerical simulation of reionization including X-ray sources and multi-frequency heating over hundreds of Mpc. Using multi-frequency radiative transfer (RT) modelling, we track the morphology of the heating and evolution of ionized regions using density perturbations and haloes obtained from a high-resolution, $N$-body simulation. The size of our simulations ($349\,$Mpc comoving on a side) is sufficient to capture the large-scale patchiness of reionization and to make statistically meaningful predictions for future 21-cm observations. We compare two source models, one with and one without X-ray sources, that otherwise use the same underlying cosmic structures. We also test the limits of validity of the common assumption that for late times the IGM temperature is much greater than the spin temperature.

The outline of the paper is as follows. In Section~\ref{sec:sims}, we present our simulations and methodology. In Section \ref{sec:theory}, we describe in detail the theory behind our generation of the 21-cm signatures. Section~\ref{sec:results} contains our results, which include the reionization
and temperature history and morphology. We also present our 21-cm maps and various statistics of the 21-cm signal. We then conclude in Section~\ref{sec:conclusions}.

The cosmological parameters we use throughout this work are ($\Omega_\Lambda$, $\Omega_\mathrm{M}$, $\Omega_\mathrm{b}$, $n$, $\sigma_\mathrm{8}$, $h$) = (0.73, 0.27, 0.044, 0.96, 0.8, 0.7), where the notation has the usual meaning and $h = \mathrm{H_0} / (100 \  \mathrm{km} \ \rm{s}^{-1} \ \mathrm{ Mpc}^{-1}) $. These values are consistent with the latest results from WMAP \citep{Komatsu2011} and Plank combined with all other available constraints \citep{Planck2015,Planck2016}.

\section{THE SIMULATIONS}
\label{sec:sims}

In this section, we present an overview of the methods used in our simulations. We start with a high-resolution, $N$-body simulation, which provides the underlying density fields and dark matter halo catalogues. We then apply ray-tracing RT to a density field that is smoothed to a lower resolution to speed up the calculations. The sources of ionizing and X-ray radiation are associated with the dark matter haloes. Below, we describe these steps in more detail.

\subsection{$N$-BODY SIMULATIONS}

The cosmic structures underlying our simulations are based on a high-resolution $N$-body simulation using the \textsc{\small CubeP$^3$M} code \citep{Harnois2013}. A 2-level particle-mesh is used to calculate the long-range gravitational forces, kernel-matched to local, direct and exact particle-particle interactions. The maximum distance between particles over which the direct force is calculated set to 4 times the mean interparticle spacing, which was found to give the optimum trade off between accuracy and computational expense. Our $N$-body simulation follows $4000^3$ particles in a $349\,Mpc$ per side volume, and the force smoothing length is set to 1/20th of the mean interparticle spacing (this $N$-body simulation was previously presented in \citet{2016MNRAS.456.3011D} and completed under the Partnership for Advanced Computing in Europe, PRACE, Tier-0 project called PRACE4LOFAR). This particle number is chosen to ensure reliable halo identification down to 10$^9$ $\rm M_\odot$ (with a minimum of 40 particles).

The linear power spectrum of the initial density fluctuations was calculated with the code \textsc{\small CAMB} \citep{Lewis2000}. Initial conditions were generated using the Zel'dovich approximation at sufficiently high redshifts (initial redshift $z_\mathrm{i}=150$) to ensure against numerical artefacts \citep{Crocce2006}.

\subsection{SOURCES}

Sources are assumed to live within dark matter haloes, which were found using the spherical overdensity algorithm with an overdensity parameter of 178 with respect to the mean density. We use a sub-grid model \citep{Ahn15a} calibrated to very high resolution simulations to add the haloes between $10^8 - 10^9\,$M$_\odot$, the rough limit for the atomic-line cooling of primordial gas to be efficient. 

For a source with halo mass, $M$, and lifetime, $t_s$, we assign a stellar ionizing photon emissivity according to
\be
\dot{N}_\gamma=g_\gamma\frac{M\Omega_{\rm b}}{\mu m_{\rm p}(10\,\rm Myr)\Omega_0},
\ee
where the proportionality coefficient $g_{\gamma}$ reflects the ionizing photon production efficiency of the stars per stellar atom, $N_\mathrm{i}$, the star formation efficiency, $f_\star$, and the escape fraction, $f_{\rm esc}$:
\be
g_\gamma=f_\star f_{\rm esc}N_{\rm i}\left(\frac{10 \; {\rm Myr}}{t_{\rm s}}\right).
\ee
\citep[][]{Haim03a,Ilie12a}.

Sources hosted by high-mass haloes (above $10^9 \ $M$_\odot$) have efficiency $g_\gamma=1.7$ and are assumed to be unaffected by radiative feedback, as their halo mass is above the Jeans mass for ionized ($\sim\!10^4$K) gas. Low-mass haloes (between 10$^8 \ $M$_\odot$ and 10$^9 \ $M$_\odot$) have a higher efficiency factor $g_\gamma=7.1$, reflecting the likely presence of more efficient Pop.~III stars or higher escape fractions \citep{2007MNRAS.376..534I}. The low-mass sources are susceptible to suppression from photoionization heating. In this work, we assume that all low-mass sources residing in ionized cells (with an ionized fraction greater than 10 per cent) are fully suppressed, i.e.,\ they produce no ionizing photons \citep{2007MNRAS.376..534I,2016MNRAS.456.3011D}. The source model for the stellar radiation is identical to the one in simulation LB2 in \citet{2016MNRAS.456.3011D} and uses the aggressive suppression model `S' defined there. However, the details of this suppression are not very significant here, since we focus on the very early stages of reionization before significant ionization develops. The stellar sources are assigned a blackbody spectrum with an effective temperature of $T_{\mathrm{eff}} = 5 \times 10^4 K$.

The X-ray sources are also assumed to reside in dark matter haloes. They are assigned a power-law spectrum with an index of $\alpha = -1.5$ in luminosity. \citet{Hickox2007} showed, using the \textit{Chandra Deep Fields (CDFs)}, that X-ray sources with a single power law spectrum would over-contribute to the observed X-ray background (XRB) if $\alpha \lsim 1$, thus requiring softer spectra for the reionization sources. 

The frequency range emitted by these sources extends from 272.08 eV to 100 times the ionization threshold of for doubly ionized helium (5441.60 eV) (\citet{Mineo2012, Mesinger2013}).  Photons with frequencies below the minimum frequency are assumed to be obscured, as suggested by observational works \citep[e.g.\ ][]{Lutivnov2005}. This value for the minimum frequency is consistent with the optical depth from high-redshift gamma-ray bursts \citep{Totani2006,Greiner2009} and is consistent with \cite{Mesinger2013}. 

The X-ray luminosity is also set to be proportional to the halo mass, since HMXBs are formed from binary systems of stars and stellar remnants. Unlike their stellar counterparts, these sources have the same efficiency factor for all active sources. Low-mass haloes that are suppressed are assumed not to produce X-ray radiation. The efficiency is parametrised as follows:
\begin{equation}
g_{\rm x} = N_{\rm x} f_\star\left(\frac{10 \;\mathrm{Myr}}{t_{\rm s}}\right)\,
\end{equation}
where $N_{\rm x}$ is the number of X-ray photons per stellar baryon. A value of $N_{\rm x}=0.2$ is roughly consistent with measurements between 0.5--8~keV for X-ray binaries in local, star-bursting galaxies \citep{Mineo2012} although the uncertainty is a factor 2 to 3 \citep[see the discussion in][]{Mesinger2013}. We take $g_\mathrm{x} = 8.6\times 10^{-2}$, which implies $f_\star\approx 0.4$ if $N_{\rm x}=0.2$. Our X-ray luminosities are, therefore, somewhat higher than in the local Universe. The total number of X-ray photons contributed from these sources over the simulation time is an order of magnitude lower than the value obtained from the CDFs for the XRB between 1 and 2 keV \citep{Hickox2007} making these sources consistent with observations. Note that the long range X-ray heating we examine here is dependent on the abundance, clustering and spectra of the X-ray sources. We leave comparisons between different X-ray source models for future work.

\subsection{RT SIMULATIONS}

The RT is based on short-characteristics ray-tracing for ionizing radiation \citep[e.g.][]{1999RMxAA..35..123R} and non-equilibrium photoionization chemistry of hydrogen and helium, using the code \textsc{\small C$^2$-Ray}, \textbf{C}onservative, \textbf{C}ausal \textbf{Ray} tracing \citep{Mellema2006}. \textsc{\small C$^2$-Ray} is explicitly photon-conserving in both space and time due to the finite-volume approach taken when calculating the photoionization rates and the time-averaged optical depths used. This quality enables time-steps much longer than the ionization time scale, which results in the method being orders-of-magnitude faster than other approaches. However, we note that including the gas heating could impose some additional constraints on the time-stepping, resulting in smaller time-steps as discussed in \cite{Lee2016}.

The basic RT method was further developed in order to accommodate
multi-frequency RT \citep{Friedrich2012}, including the effects of helium, secondary ionizations from electrons, multi-frequency photoionization and detailed heating through full on-the-spot approximation, in order to correctly model the effects of hard radiation. Frequency bin-integrated rates are used for the photoionization and photoionization heating rates with three bands: one for the frequency range in which only Hydrogen may be ionized (i.e. from the first ionization level of Hydrogen to the first of Helium), a second for photons which can ionize both hydrogen and helium once and a third for photons that can ionize all species. These are sub-divided into 1, 26, and 20 sub-bins respectively. The rates values are pre-calculated and stored in look up tables as functions of the optical depths at the ionization thresholds. The convergence of the number of sub-bins in bands 2 and 3 have been tested in \citet{Friedrich2012}. They concluded 10 and 11 sub-bins for bands 2 and 3 respectively produced sufficiently converged results. \textsc{\small C$^2$-Ray} has been tested extensively against existing exact solutions \citep{Mellema2006}, numerous other numerical codes within code
comparison projects \citep{Iliev2006b,Iliev2010}, and against
\textsc{\small CLOUDY} \citep{Friedrich2012}.

In this work, we present two simulations: one in which the haloes contain both HMXB \& stellar sources, and one which only considers stellar sources. The stellar component and underlying cosmic structures are identical in both simulations. The density is smoothed onto an RT grid of size $250^3$. These simulations were performed under the PRACE Tier-0 projects PRACE4LOFAR and Multi-scale Reionization.

\section{The 21-CM SIGNAL}
\label{sec:theory}

In this section, we discuss the method of extracting the 21-cm signal from our simulation outputs.

\subsection{THE DIFFERENTIAL BRIGHTNESS TEMPERATURE}
\label{sec:dbtTheory}

Observations aim to detect the redshifted 21-cm signal caused by the hyperfine transition from the triplet to the singlet ground state of the neutral hydrogen present during reionization. This signal is dictated by the density of neutral hydrogen atoms and the ratio of hydrogen atoms in the triplet and singlet states, quantified by $T_{\mathrm{S}}$:
\begin{equation}
\frac{N_1}{N_0}=\frac{g_1}{g_0}\exp\left(-\frac{T_\star}{T_{\rm S}}\right).
\end{equation}
Here $T_\star=\frac{h\nu_{10}}{k}=0.0681~$K is the temperature corresponding to of the 21-cm transition energy, and $g_{1,0}$ are the statistical weights of the triplet and singlet states, respectively. 

For the 21-cm signal to be visible against the CMB, the spin temperature needs to decouple from it, since the two start in equilibrium at high redshift. The two mechanisms that can do this \citep{Field1958}. Firstly, collisions with other atoms and free electrons do so by exciting electrons from the singlet to the triplet state. This mechanism is only effective for sufficiently dense gas, i.e. in very dense filaments and haloes or at very high redshifts. Secondly, the electrons can be excited to the triplet state through the Wouthuysen-Field (WF) effect when absorbing a \Lya\ photon. The spin temperature can then be expressed as follows \citep{Field1958}:
\begin{equation}
T_{\mathrm{S}} = \frac{T_{\mathrm{CMB}} + x_\mathrm{\alpha} T_\mathrm{c} + x_\mathrm{c} T_\mathrm{k}}{1 + x_\mathrm{\alpha} + x_\mathrm{c}},
\end{equation}
where $T_\mathrm{c}$ is the \Lya\ colour temperature, $x_\mathrm{\alpha}$ is the \Lya\ coupling constant, $T_\mathrm{k}$ is the gas kinetic temperature, and $x_\mathrm{c}$ is the collisional coupling constant. Throughout this paper, we assume that \Lya\ radiation is in the WF-effect-saturated-regime (in which case $x_\mathrm{\alpha} \gg x_\mathrm{c}$), and the colour temperature is equal to the kinetic temperature ($T_\mathrm{c} = T_\mathrm{k}$); hence, $T_\mathrm{S} = T_\mathrm{k}$. Since early sources produce copious amounts of soft-UV photons, this approximation tends to hold throughout most of the evolution, except for the earliest times \citep[e.g.][]{2003ApJ...596....1C}.

The 21-cm signal itself is usually defined in terms of the differential brightness temperature with respect to the CMB:
\begin{equation}
\delta T_{\mathrm{b}} = \left(1 - \frac{T_{\mathrm{CMB}}}{T_{\mathrm{S}}}\right) \ \frac{3 \lambda_0^3 A_{10} T_\star n_{\mathrm{HI}}(z)}{32 \pi T_{\mathrm{S}} H(z) (1+z)} 
\label{dbt},
\end{equation}
where $\lambda_0=21.1$~cm is the line rest-frame wavelength, $A_{10}=2.85\times10^{-15}\,\rm s^{-1}$ is the Einstein A-coefficient for spontaneous emission from the triplet to singlet state, and $n_{\mathrm{HI}}$ is the density of neutral hydrogen. Thus, $\delta T_{\mathrm{b}}$ could be seen either in absorption or emission, depending on the $T_{\rm s}$ relative to the CMB. We are particularly interested in the timing and character of the transition between absorption and emission. Predicted 21-cm maps (smoothed to the resolution of observations) and their statistical measures will be the only way to connect theories of galaxy formation to future observations of 21-cm radiation. 

When calculating the 21-cm signal, many studies assume the IGM gas to have reached temperature saturation, i.e. to be heated to temperatures well above the CMB, $T_\mathrm{K} \gg T_{\mathrm{CMB}} $. We refer to this as the high-temperature limit, high-$T_\mathrm{K}$ limit. While this approximation may hold during the later stages of reionization, it certainly breaks down at early times. Where appropriate, we show the high-$T$$_\mathrm{K}$ limit results for reference.

\subsection{TEMPERATURE OF THE NEUTRAL IGM IN PARTIALLY IONIzED REGIONS}
\label{sec:corrections}
H~II regions can have sizes smaller than our cell resolution, particularly for individual weak sources, and therefore be unresolved in our simulations. The cells containing such ionized regions will appear partially ionized in the simulation, with a temperature that is averaged between the hot, ionized gas phase and the colder, neutral one. At 21-cm, using the average cell T would yield a signal in emission where it should appear in absorption. 
In order to correct for such unphysical behaviour, we have adopted an algorithm to locate such cells and calculate $\delta T_{\rm b}$ appropriately, as follows. Since this does not affect any of the physical quantities produced by the code, it can be performed as a post-processing step.

\textit{Finding and marking the cells requiring special treatment:} Since the softer stellar spectra do not produce photons that can penetrate into the IGM (the typical mean free paths are of order kpc), the cells potentially requiring correction in the stellar-only simulation are identified as those with $T > T_\mathrm{ad}$ and $x>x_{\rm in}$, where $T_\mathrm{ad}$ is the mean adiabatic gas temperature of the universe and $x_{\rm in}$ is the initial ionized fraction. In the HMXB simulation the cells that might need correction are in the same locations as in the stellar-only simulation since the ionizing sources are identical between the two simulations and the co-located HMXBs do not contribute enough ionizing photons to significantly grow the primarily stellar radiation-driven ionized regions (which we have tested using high resolution simulations and analytic estimates).

\textit{Calculating the temperature of H~II regions in the stellar-only simulation:} The softer stellar spectra yield sharp H~II region boundaries, separating them from the cold, adiabatically cooling IGM (due to the lack of shock heating in our simulations, the treatment of which we leave for future work). We thus assume that the temperature of the neutral gas is that adiabatic temperature, $T_\mathrm{HI,s}=T_\mathrm{ad}$. The temperature of the H~II regions in each marked cell is thus calculated using:
\begin{equation}
T_\mathrm{HII,s} = \frac{T_\mathrm{c,s} - T_\mathrm{ad} (1-x)}{x},
~\label{thii}
\end{equation}
where $T_\mathrm{HII,s}$ is the temperature of the H~II region, $T_\mathrm{c,s}$ is the cell-average temperature given by \textsc{\small C$^2$-Ray}, $T_\mathrm{ad}$ is the adiabatic temperature of the universe and $x$ is the volume weighted ionized fraction of H.

\textit{Calculating $\mathrm{T_{HI}}$ from HMXB simulation:} The temperature of the neutral IGM in each marked cell in the HMXBs simulation is calculated using:
\begin{equation}
T_\mathrm{HI,x} = \frac{T_\mathrm{c,x} - T_\mathrm{HII,s} x}{1-x},
\end{equation}
where 
$T_\mathrm{c,x}$ is the temperature from $\small $C$^2$-RAY from the HMXB simulation and $T_\mathrm{HII,s}$ is from Equation ~\ref{thii}. Here, we assume that the temperature of the H~II regions is the same between the two simulations, since the local heating in the H~II region is strongly dominated by the stellar emission. This similarity was verified by high-resolution tests and analytical estimates, which showed that the additional X-ray heating is negligible. 

In some cases, the LMACHs are suppressed in the HMXB case, but not the stellar-only case due to a marginally higher ionized fraction. These very rare cells are taken to be the average temperature of their neighbours. In summary, we use the temperatures of the neutral gas in each cell ($T_\mathrm{HI,s}$ and $T_\mathrm{HI,x}$, respectively for the two simulations), as calculated above to derive the 21-cm $\delta T_{\rm b}$.

\section{RESULTS}
\label{sec:results}

\subsection{REIONIZATION AND THERMAL HISTORIES}
\label{sec:reion_hist}
\begin{figure}  
\includegraphics[width=3.6in]{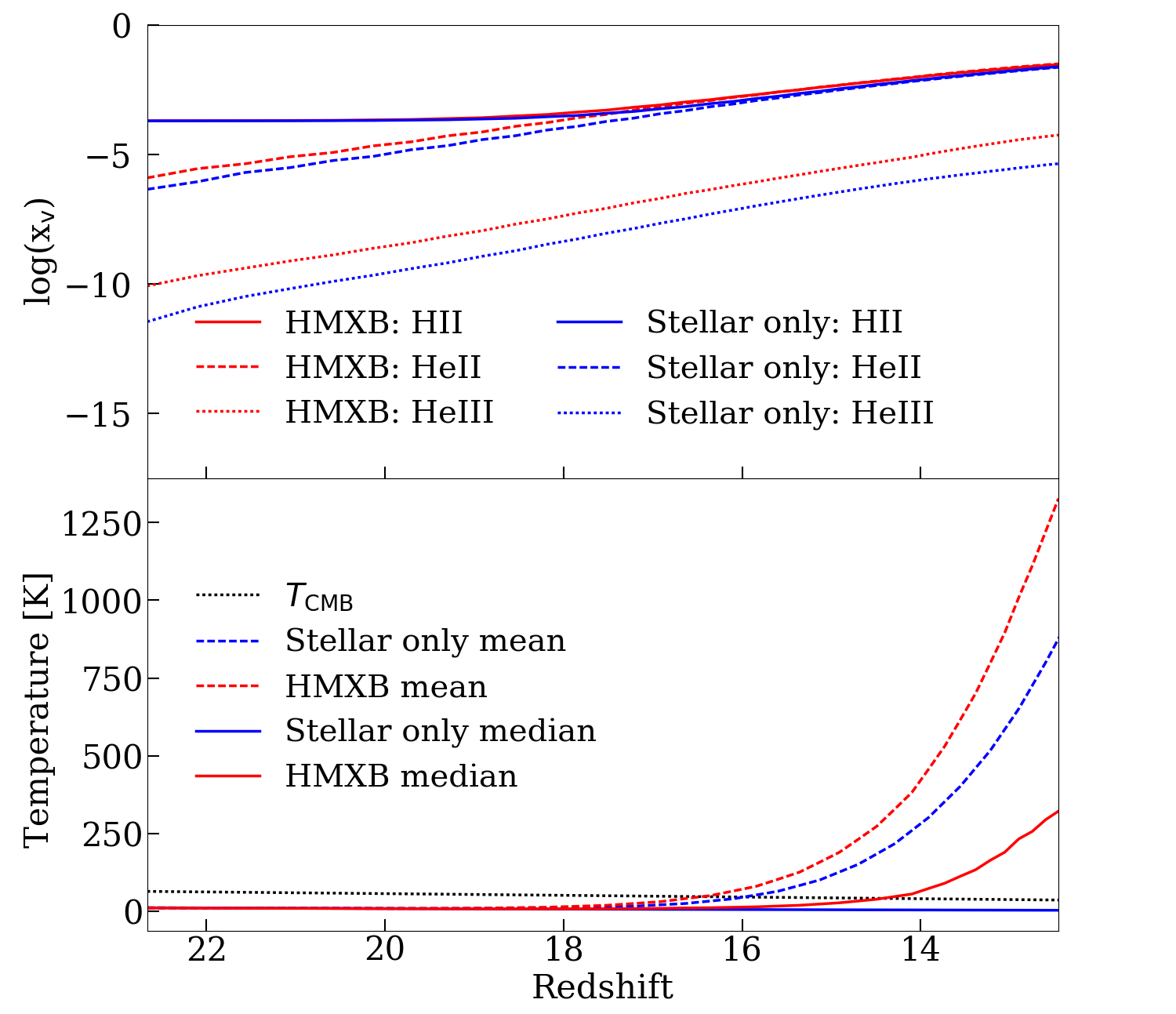}
\caption{(top) The mean ionized fraction by volume of each species: HII - solid lines, HeII - dashed lines and HeIII - dotted lines. (bottom) The volume-weighted mean temperature - dashed lines, median temperature - solid lines and T$_{\rm CMB}$ - dotted line. In both plots the HMXB case is shown in red and the stellar-only case is shown in blue. \label{fig:meanhist}}
\end{figure}

The 21-cm signal is affected by the thermal and ionization histories of the IGM. In the upper panel of Fig.~\ref{fig:meanhist}, we show the mean volume-weighted ionized fraction evolution for the species present in our simulations: H~II (solid lines), He~II (dashed lines), and He~III (dotted lines) for the HMXB case (red lines) and the stallar only case (blue lines). The ionization of H~I and He~I is largely driven by the hard-UV photons of stars, which are more abundant than X-ray photons. The effect of the latter is largely limited to increasing the He~III abundance by about an order of magnitude (while remaining low) due to their high energy per photon. 

In the lower panel, volume-weighted mean (dashed lines) and the median (solid lines) temperatures are shown for the HMXB case (red lines) and the stellar-only case (blue lines). For the HMXB case, the mean temperatures are increased modestly, surpassing $T_{\mathrm{CMB}}$  earlier than in the stellar-only case, with both occurring around $z\sim16$. The differences between the mean $T_{\rm K}$ of the two cases (with and without X-rays) grow at later times, rising above 50~per~cent for $z<13$. As the 21-cm signal probes neutral hydrogen, the median temperature of the IGM is more relevant, since the mean is skewed towards higher valued by the hot, ionized regions. In the presence of X-rays, the median surpasses $T_{\mathrm{CMB}}$ just before $z=14$; while in their absence the neutral IGM remains cold. 

The only previous full numerical simulations to take into account X-ray sources are from \citet{Baek2010}\footnote{The simulations in \citet{Xu2014} and \citet{Ahn2015} are focused on a single X-ray source in a zoomed region of a cosmological volume, so are not directly comparable to our results here.}. However, the mass resolution of these simulations is approximately 600 times lower than ours, and the volume is 15 times smaller. As a consequence, their first sources of any type of radiation appear around $z\approx 14$, approximately when the neutral IGM in our simulation has already been globally heated to well above the $T_{\rm CMB}$. In other words, their simulations describe a scenario in which X-ray sources appear very late, are relatively rare and bright, and are coincident with substantial stellar emission. This situation is very different from our case in which the first X-ray sources appear at $z\approx 23$ and large numbers of relatively faint sources heat the neutral IGM well before any substantial reionization. We will, therefore, not further compare the details of our results to those of \citet{Baek2010}.

\begin{figure} 
\includegraphics[width=0.475\textwidth]{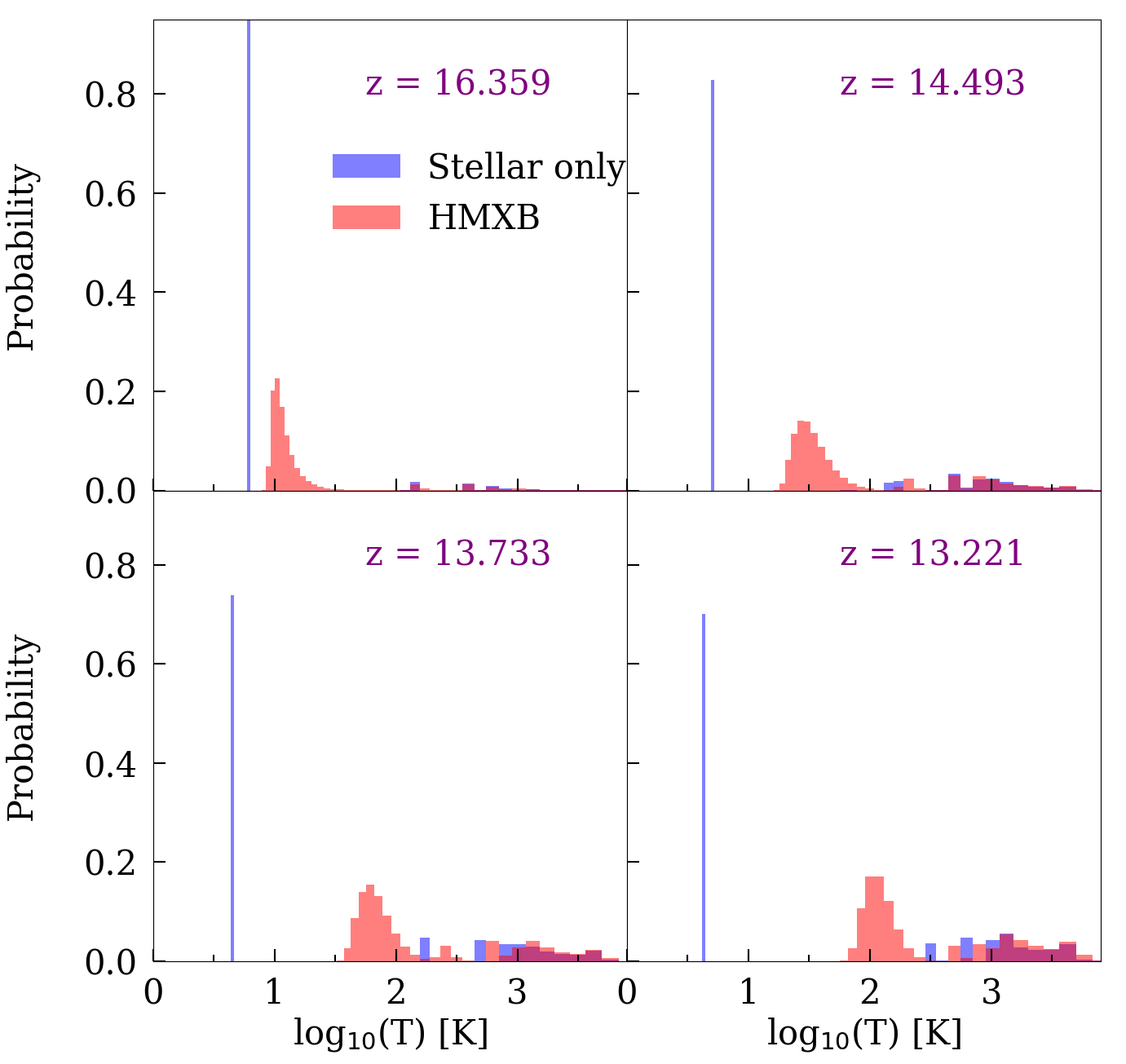}
\caption{Histograms of the temperature at the full simulation resolution
for the HMXB (red) and stellar-only (blue) cases for several illustrative redshifts. 
\label{fig:hist}}
\end{figure}

More detailed information about the temperature distributions is obtained from the corresponding probabily distributions functions (PDFs), shown in Fig.~\ref{fig:hist}. Again, the HMXB case is shown in red and the stellar-only case in blue. These PDFs were generated from the coeval simulation cubes using 100 bins and normalised to have a total area of one. The stellar-only distributions are clearly bimodal, with a few hot, partially ionized regions and the majority of cells remaining very cold. This behaviour is expected, given the very short mean free path of the ionizing photons in this case, which yields sharp ionization fronts. In contrast, when X-rays are present, their long mean free paths lead to gas heating spreading quickly and widely, with all cells being affected. The distribution is strongly peaked, relatively wide, and gradually moves towards higher temperatures, with typical values above 100~K by $z=13.2$. Our thermal history is similar to that of Case A (Pop.~II stars) in \citet{Pritchard2007} and case `$\log\zeta_{\rm X}=55$' in \citet{2015MNRAS.454.1416W}. These studies do not provide temperature PDFs or median values. 

\begin{figure*} 
\centering
\includegraphics[width=0.8\textwidth]{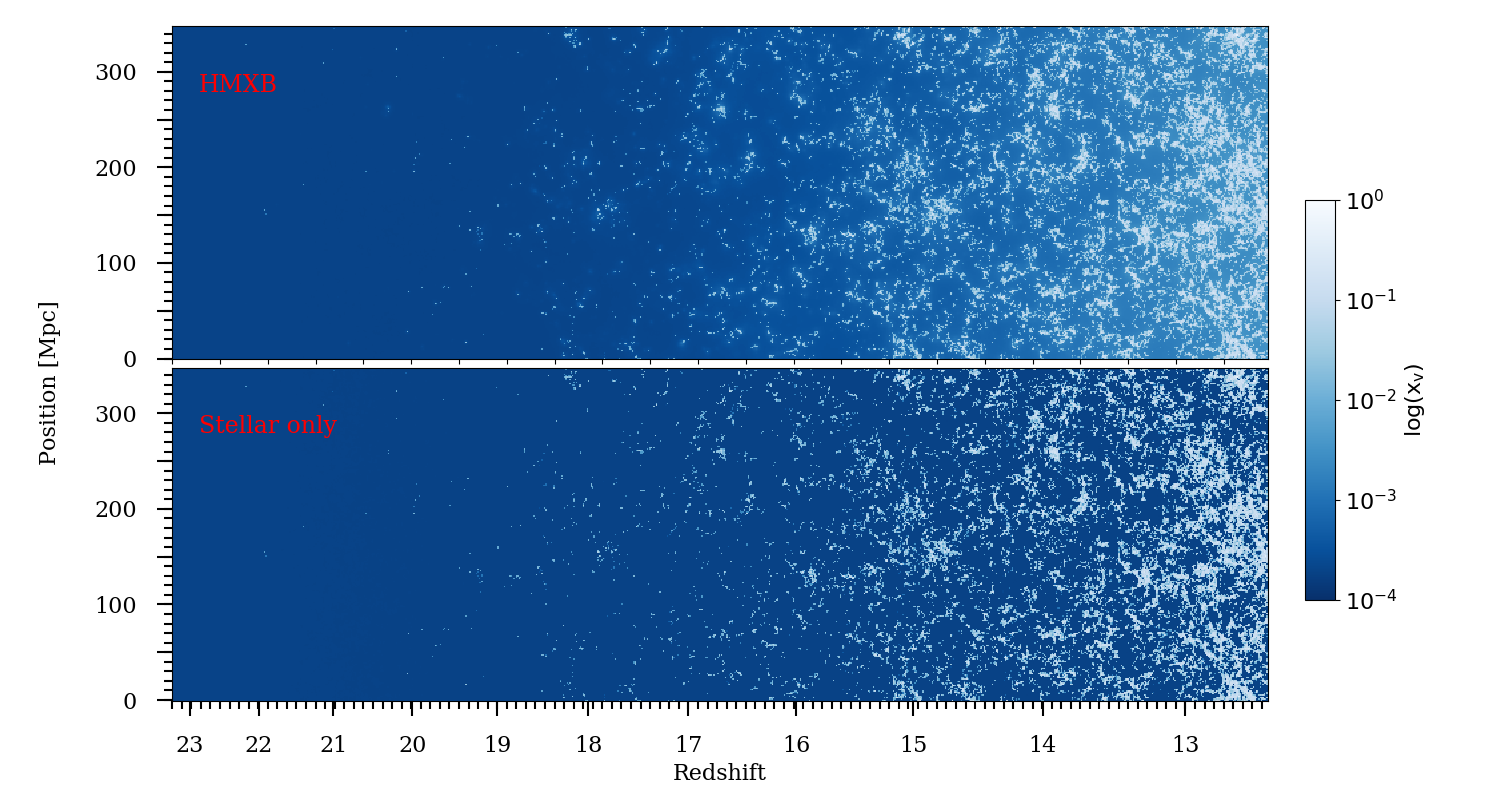}
\includegraphics[width=0.8\textwidth]{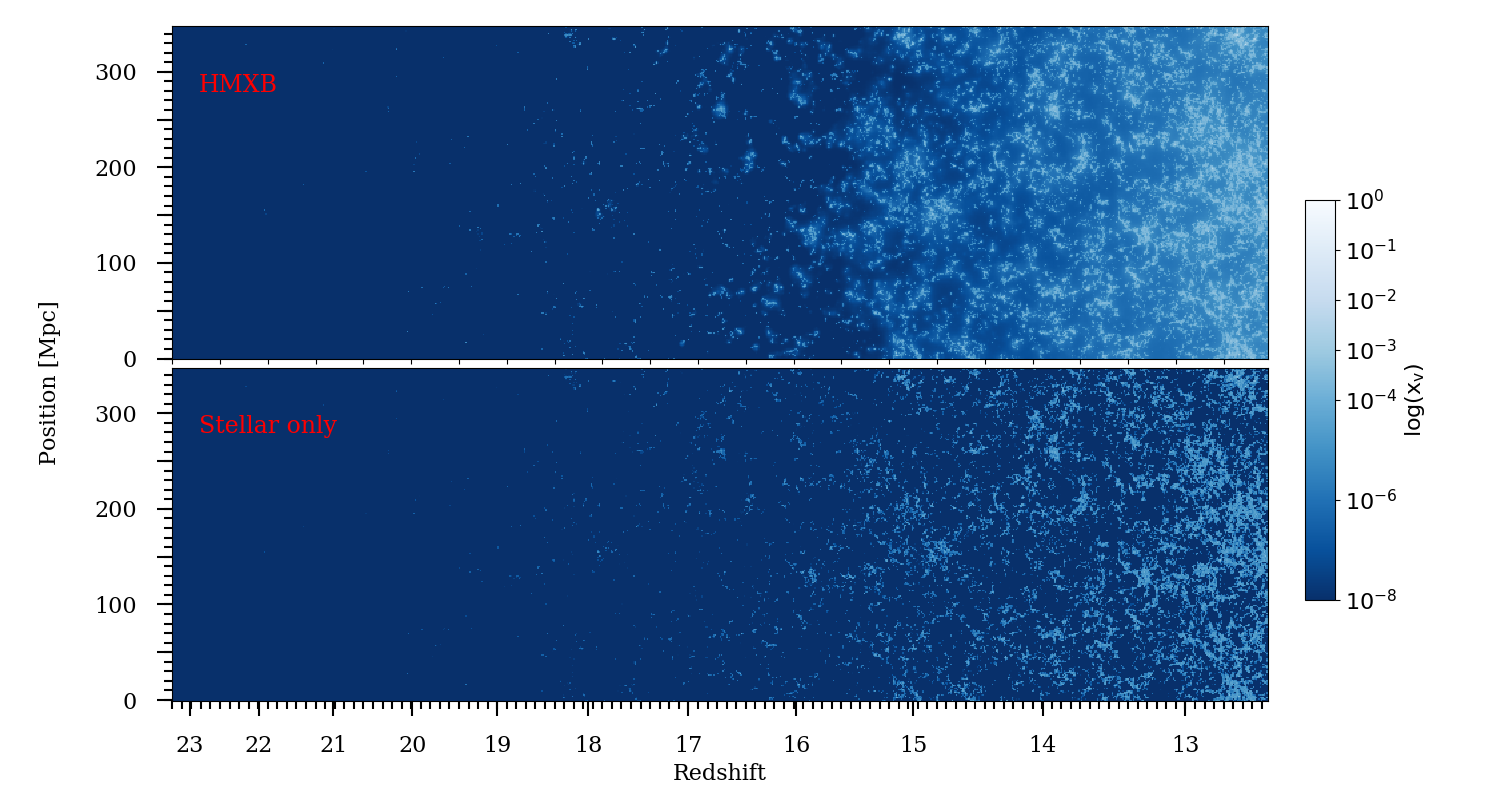}
\caption{The position-redshift lightcone images of the ionized volume fraction
of hydrogen (top two panels) and He~III (bottom two panels). Shown are the HMXB case (top panel in each pair), and the stellar-only case (lower panel in each pair).}
\label{fig:lightcones:xfrac}
\end{figure*}

The lightcones (spatial-redshift/frequency slices) provide a visual representation of the quantities discussed above, including spatial variations and evolution over time (Fig.~\ref{fig:lightcones:xfrac}). These lightcones are constructed by taking a cross section of the simulation volume along the line of sight and continually interpolating in time the relevant quantity using the spatial periodicity of our cosmological volume. 

In Fig.~\ref{fig:lightcones:xfrac} (top panels), we show the hydrogen ionization lightcone. As expected based on the very similar mean ionization fractions in the two simulations, the morphology of hydrogen ionization is broadly similar. However, the hard photons, which penetrate deep into the neutral regions produce a low-level, but widespread ionization of the IGM and `fuzzier', less clearly defined H~II regions when X-ray sources are present. Given their similar ionization potentials, the first ionization of helium (not shown here) closely follows that of hydrogen. Alternatively, the second helium ionization potential is sufficiently higher to result in significant differences between models (Fig.~\ref{fig:lightcones:xfrac}, bottom panels). The 50,000~K blackbody stellar spectra produce very few photons able to fully ionize a helium atom; thus, any He~III produced is concentrated in the immediate surroundings of the ionizing sources. However, the X-rays are very efficient in fully ionizing helium, producing widespread ionization (albeit still at a relatively low level). This ionization is also quite patchy on large scales, especially at early times. The exact morphology depends on the spectra, abundance, and clustering of the X-ray sources. 

The lightcones in Fig.~\ref{fig:lightcones:temp} (top panels) show the spatial geometry and evolution of the IGM heating. The soft, stellar radiation in both models ionize and heat the immediate environments of the sources to $T\sim10^4$~K, seen as dark regions, with the majority of the IGM remaining completely cold. The X-ray radiation propagates much further, starting to heat the gas throughout. Large, considerably hotter regions (black in the image) develop, quickly reaching tens of Mpc across before $z\sim15$. These regions gradually expand and merge, resulting in thorough heating to hundreds of degrees by $z\sim14$; though, large cold regions still remain present down to $z\sim13.5$. At early times, the heating from X-rays increases the inhomogeneity when X-rays are present, but the temperature distribution eventually becomes more homogeneous at later times. 

\subsection{21-CM DIFFERENTIAL BRIGHTNESS TEMPERATURE} 
\label{sec:21-cm_maps}

\begin{figure*} 
\centering
\includegraphics[width=0.8\textwidth]{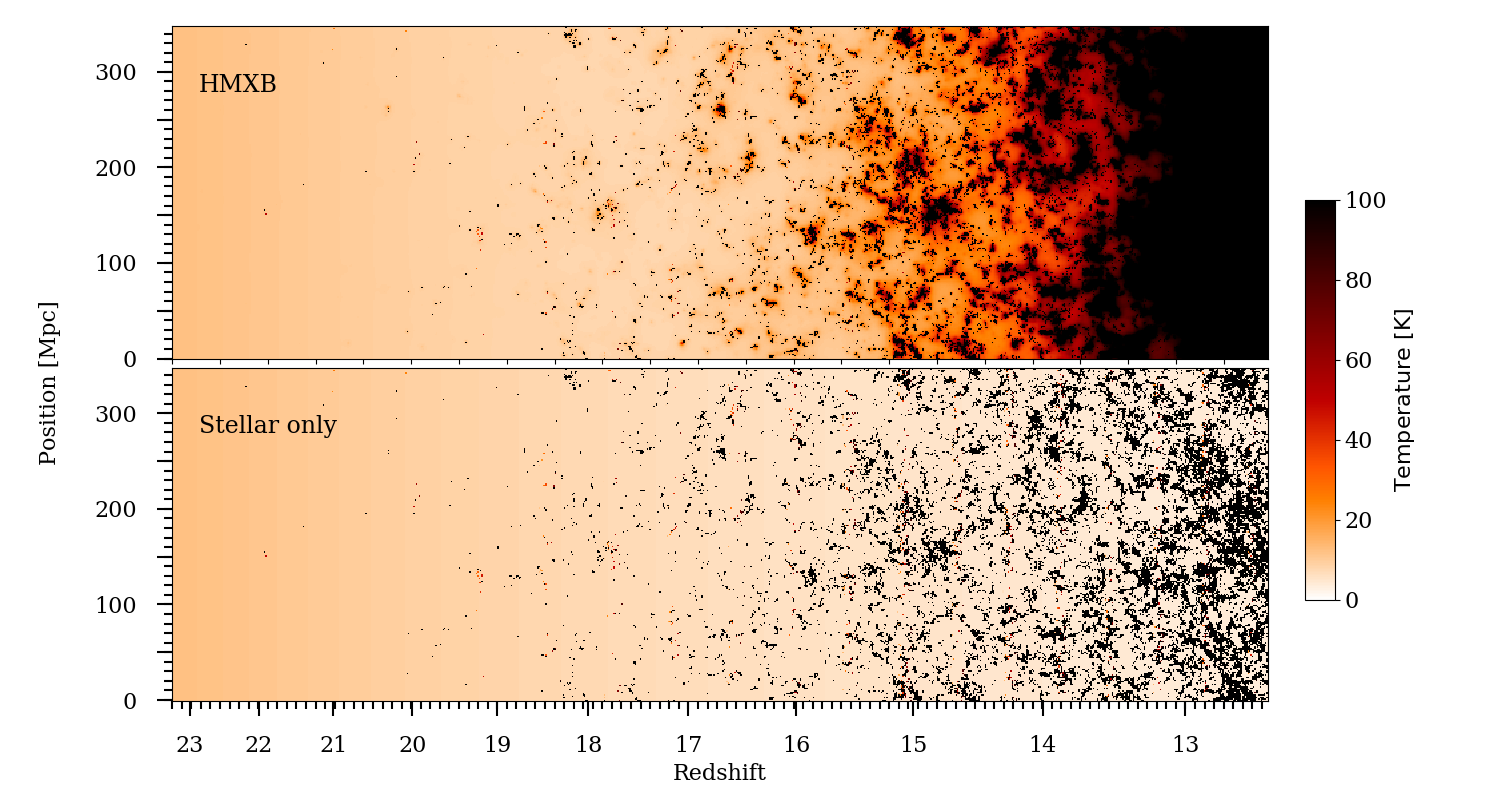}
\includegraphics[width=0.8\textwidth]{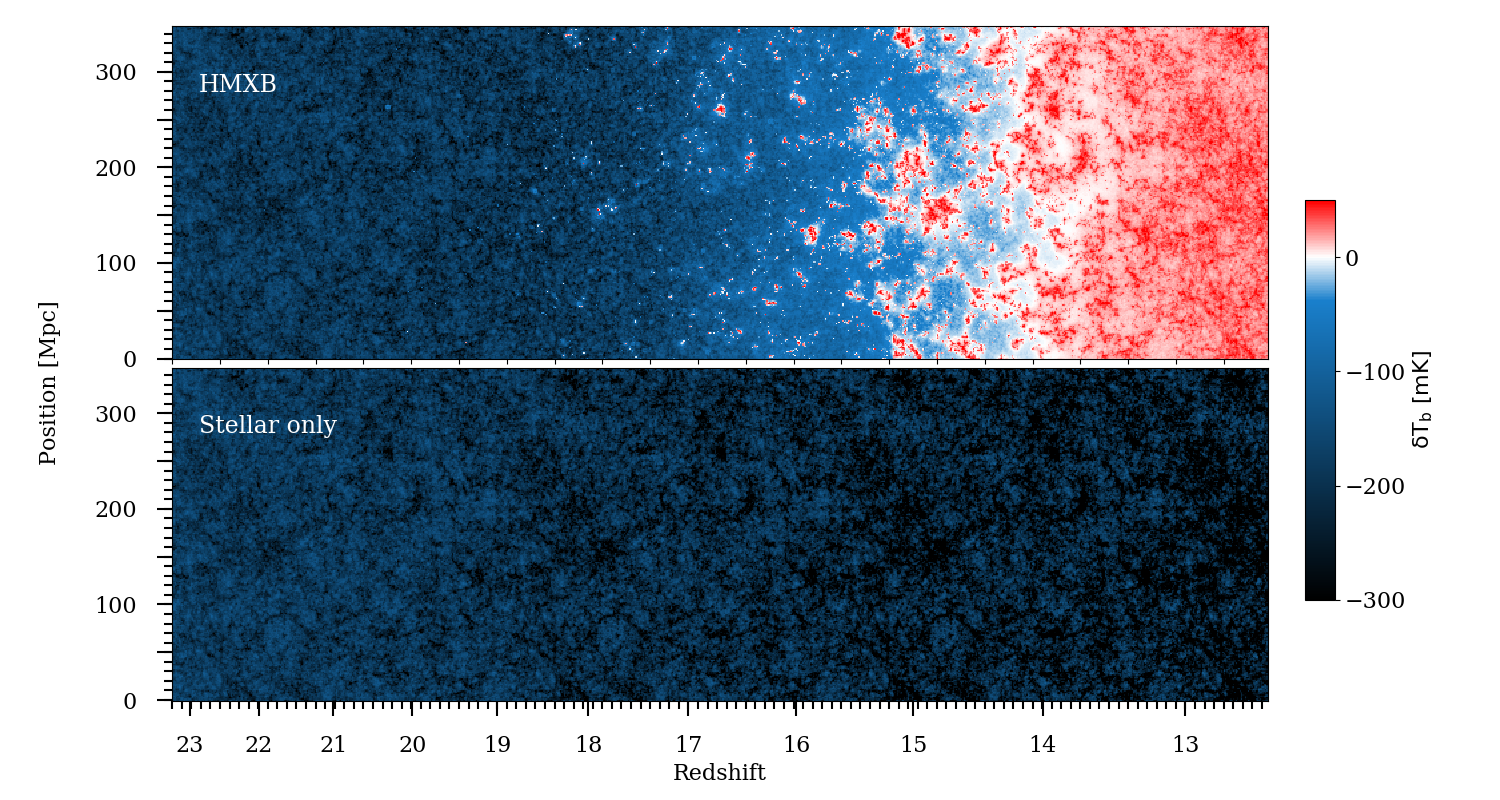}
\caption{The position-redshift lightcone images of the IGM gas temperature (top two panels) and 21-cm differential brightness temperature (bottom two panels). Shown are the HMXB case (top panel in each pair), and the stellar-only case (lower panel in each pair).}
\label{fig:lightcones:temp}
\end{figure*}

We are primarily interested in the directly observable quantity of this epoch, the 21-cm $\delta T_{\rm b}$. As discussed in Section~\ref{sec:theory}, $\delta T_{\rm b}$ depends on the density, ionization, and temperature fields. In Fig.~\ref{fig:lightcones:temp} (bottom panels), we show the $\delta T_\mathrm{b}$ lightcones corresponding to the same cross section through the position-redshift image cube as in Fig.~\ref{fig:lightcones:xfrac} and Fig.~\ref{fig:lightcones:temp} (top panels). The morphology of the $\delta T_\mathrm{b}$ fluctuations is closely related to that of the heating, demonstrating the importance of temperature variations, especially during the early stages of the EoR. Long-range X-ray heating produces a gradual, extended transition from absorption into emission. Large-scale fluctuations are significant throughout, and the first emission regions appear part way through the simulation at $z=18$, after enough X-rays have penetrated into the IGM near the ionizing sources in order to heat it. After this these bubbles grow quickly, with some reaching tens of Mpc in size by $z\sim15.5$. Only after $z\sim13.8$ have all large regions of 21-cm absorption disappeared.

Comparing to \citet{Mesinger2013}, the results of our simulation appear to be closest to their case with X-ray efficiency $f_{\rm x}=1$. They also find an extended transition from absorption to emission starting at $z\sim20$, which completes around $z\sim14$. However, by that time, the hydrogen ionization fraction is around 10~per~cent, which is substantially higher than our case. The majority of the difference is likely due to the fact that their sources are more efficient, as indicated by a completion of reionization by $z_{\rm reion}\sim8$ in their case versus $z_{\rm reion}<6.5$ for our source model \citep{2016MNRAS.456.3011D}.

In the stellar-only case, the signal remains in absorption throughout the simulation as there are no photons with long enough mean free paths to penetrate and heat the neutral IGM. As a result the IGM simply cools adiabatically as the universe expands\footnote{There is also some Compton heating due to CMB scattering, which we do include in our simulation. This heating is inefficient at this point due to the low density of the IGM.}. In reality the temperature of the IGM would also be impacted by shock heating, however, this has not been included in our simulations and is left for future work.

\begin{figure*} 
\centering
\makebox[\textwidth][c]{\includegraphics[width=1.0\textwidth]{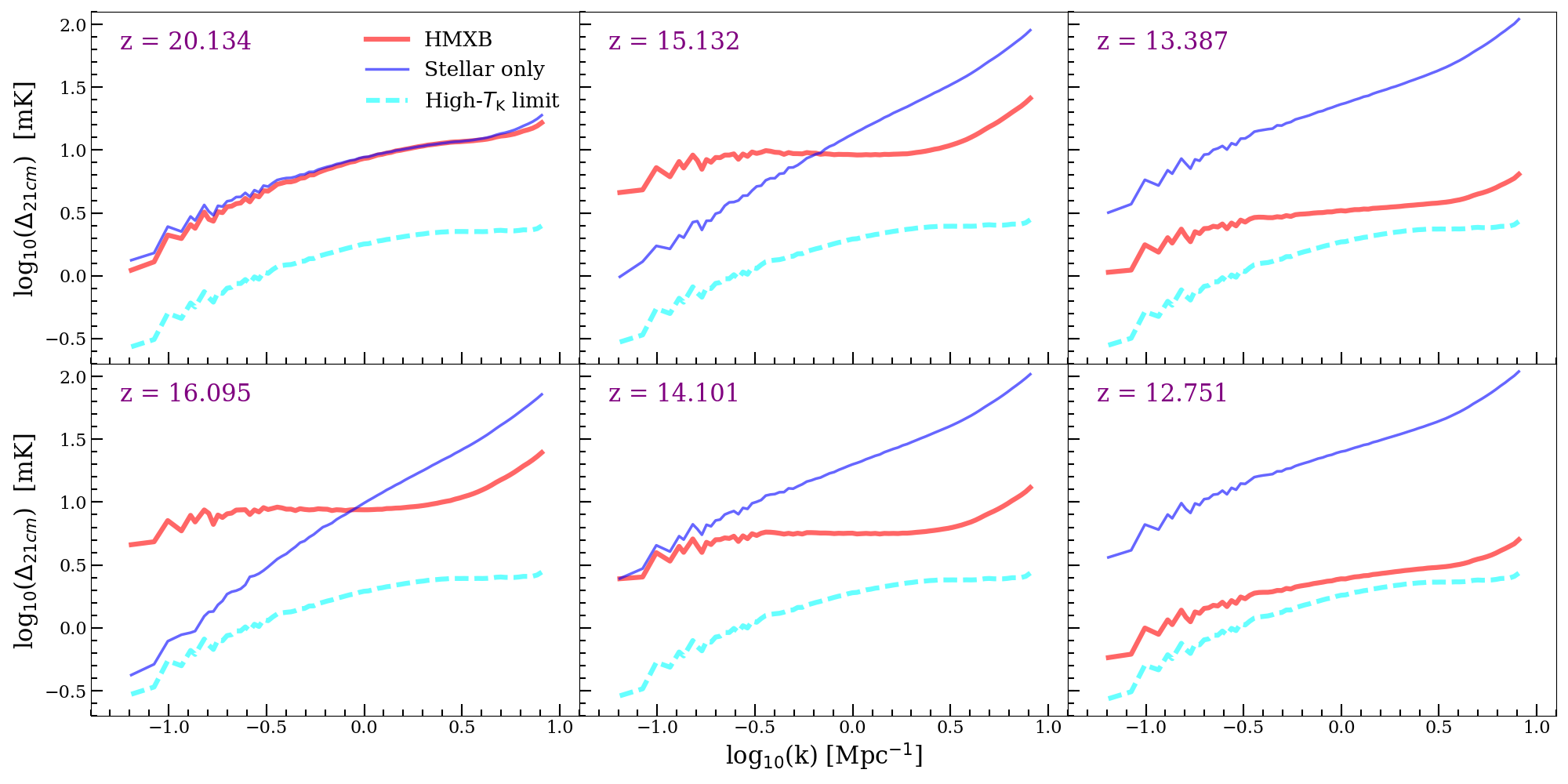}}
\caption{The 21-cm power spectra from our simulations at several key stages of the evolution with the high-$T$$_\mathrm{K}$ limit results for reference. The high-$T$$_\mathrm{K}$ limit is shown in yellow and as before the results from the HMXB case are shown in red and the stellar-only case in blue.}
\label{fig:powerspectra}
\end{figure*}

In Fig. \ref{fig:powerspectra}, we show the power spectra of $\delta T_\mathrm{b}$ at several key redshifts. During the early evolution (shown is $z=20.134$), only a modest amount of heating of the IGM has yet occurred in the X-ray model. Thus, the large-scale 21-cm fluctuations are dominated by the density variations, which are the same in the two cases. Therefore, the 21-cm power spectra are almost identical at this stage, with power suppressed slightly on all scales in the HMXB case. The high-$T_\mathrm{K}$ limit results in significantly lower fluctuations, reflecting the lower average $\delta T_\mathrm{b}$ in the emission regime compared to the absorption or mixed regimes.

As the evolution progresses and with X-ray sources contributing to the long-range heating, the 21-cm power is significantly boosted on large scales and a well-defined, if broad, peak develops around  scale of $\sim\!43$~Mpc ($z=15-16$). The small-scale power decreases due to the stronger heating in the vicinity of sources, which brings the temperature contrast with the CMB down and closer to the high-$T$$_\mathrm{K}$ limit. The overall fluctuations peak at $\sim14~$mK around $z\sim15$ in the HMXB model. \citet{Pritchard2007} find a heating peak that is in agreement with our result, at similar scales ($k\sim0.14\,\rm Mpc^{-1}$) with an amplitude of$\Delta_{\rm 21cm}\sim 11.5$mK or $\Delta_{\rm 21cm}\sim20$~mK depending on the source model used. Results from \citet{2014MNRAS.443..678P} are also in rough agreement with our own, with a power spectra peak at a similar scale (44 Mpc) and at similar redshift ($z\sim15-16$). Their peak value is also in agreement at $\Delta_{\rm 21cm}\sim14\,$mK.

Over time, as emission patches develop and the overall IGM is gradually heated ($z=12-14$), the power spectra slowly approach the high-$T_\mathrm{K}$ limit, but does not reach it even by the end of our simulation. At that point ($z=12.7$), the IGM has been heated well above $T_{\mathrm{CMB}}$ throughout and therefore is in 21-cm emission everywhere. However, the neutral IGM temperature remain at only a few hundred degrees, is still considerably spatially inhomogeneous, and so is not yet fully in the high-$T$$_\mathrm{K}$ limit, where the $\delta T_{\rm b}$ becomes independent of the actual gas temperature value. 

\begin{figure} 
\includegraphics[width=0.45\textwidth]{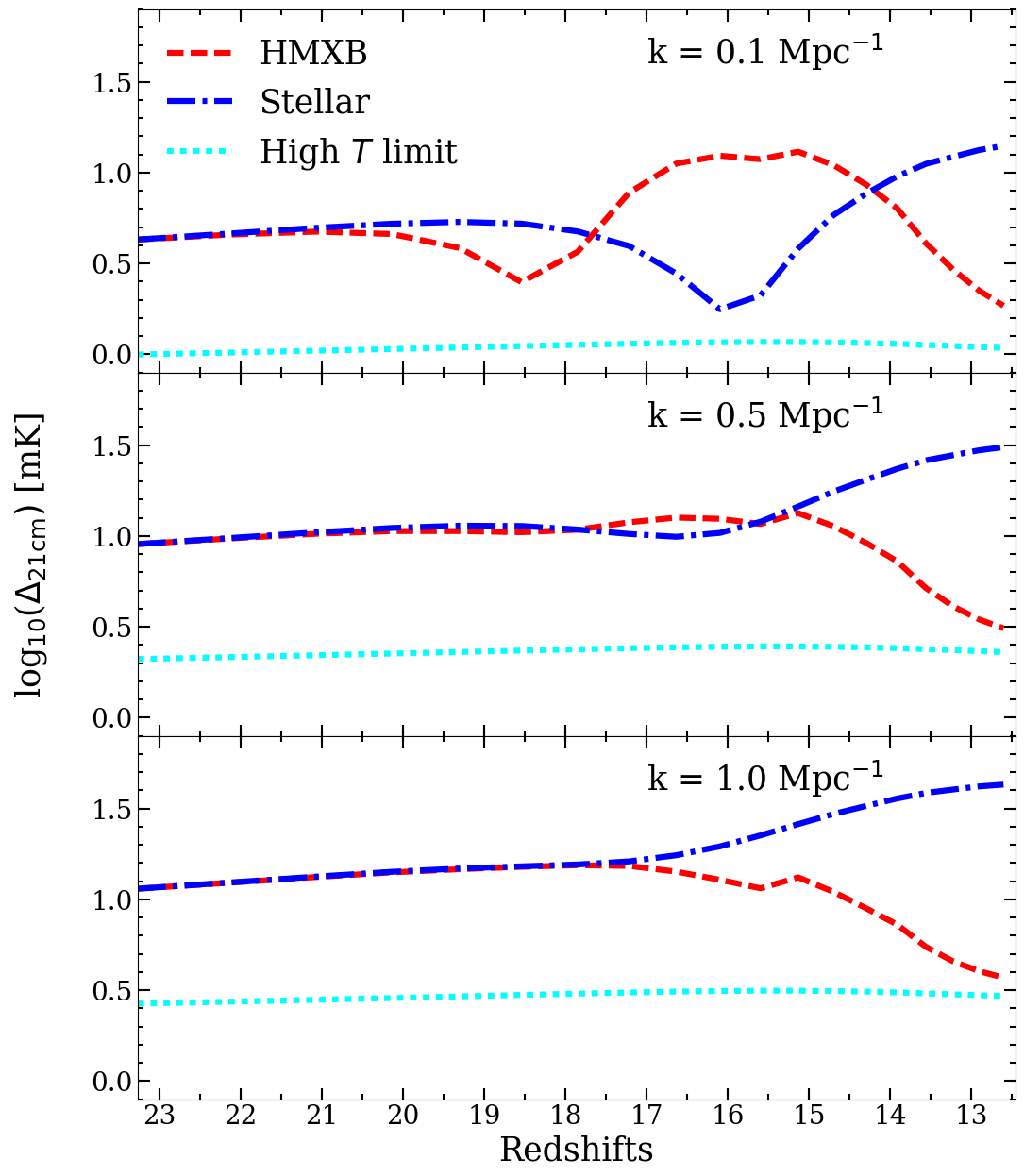}
\caption{The evolution of the 21-cm power spectra modes at $k=0.1,0.5$ and $1\,\rm Mpc^{-1}$ for the two simulations and high-$T$$_\mathrm{K}$ limit for reference, as labelled.\label{fig:kplot}}
\end{figure}
 
The evolution is markedly different in the stellar-only case. The 21-cm fluctuations here are dominated by density fluctuations as the stars do not produce many photons able to penetrate into and heat the neutral cold IGM. Consequently, the shape of the power spectrum remains almost a power law. Throughout the simulation cosmic structures continue to form so the amplitude of these fluctuations gradually increases over time. In the stellar-only case, 21-cm power on all scales is far higher at late times compared to that  in the HMXB case since the mean temperature in the latter case approaches and then surpasses $T_{\mathrm{CMB}}$ and the IGM remains very cold in the former. 

The evolution of several particular $k$-modes ($k=0.1, 0.5$, and $1\,\rm Mpc^{-1}$) is shown in Fig.~\ref{fig:kplot}. With increasing $k$, the 21-cm fluctuations deviate from the density fluctuations later due to more advanced structure formation at small scales. With X-rays present, there is a peak from heating at all scales considered here, occurring between $z\sim15 - 17$. The peak becomes wider and more pronounced at larger scales, which matches the typical 21-cm fluctuations scale due to inhomogeneous heating (on the order of tens of Mpc). At larger scales, X-ray heating from the first sources removes the colder regions close to these sources, resulting in an initial dip in the power. Subsequently, the power rises as the X-ray heating extends inhomogeneously into the IGM. Both of these features are present, but less pronounced, at the intermediate scale (k=0.1\,Mpc$^{-1}$). In the absence of X-rays, the evolution is markedly different. At later times when the IGM X-ray heating has become homogeneous, the corresponding result for the stellar-only case has higher power on all scales due to the IGM remaining cold. In the high-$T$$_\mathrm{K}$ limit, the fluctuations are much lower and mostly flat at all scales due to the lack of cold IGM, which results in lower amplitude signal driven initially by the density fluctuations only. 

The evolution of the $k=0.1\,\rm Mpc^{-1}$ mode is roughly in agreement with \citet{Mesinger2013}, however, their peak occurs earlier than our results, again likely due to the higher assumed efficiency of their sources. Probably for the same reason, \citet{Pritchard2007} and \citet{Fialkov2014} find the peak to occur later. Beyond the timing of the peak, the amplitude is in rough agreement with in the semi-numerical results. \citet{Mesinger2013}, \citet{Pritchard2007}, and \citet{Santos2010} find a marginally higher peak value of 20~mK, while \citet{Fialkov2014} find a lower one of 7~mK.

In Fig.~\ref{fig:dbtmaps}, we show maps of the mean-subtracted $\delta T_{\rm b}$ $(\delta T_\mathrm{b}-\overline{\delta T_\mathrm{b}}$) at several key epochs of the redshifted 21-cm evolution, smoothed with a Gaussian beam that is roughly twice as broad as the what can be achieved by the core of SKA1-Low, which will have maximum baselines of around 2~km. We averaged over a frequency bandwidth that corresponds to the same spatial scale as the Gaussian beam. To mimic the lack of sensitivity at large scales, which interferometers have due to the existence of a minimum distance between their elements, we also subtracted the mean value from the images. We do not include instrument noise and calibration effects in these maps. However, at this resolution, SKA1-Low is expected to have a noise level of $\sim 10$~mK per resolution element for 1000 hours of integration \citep{2015aska.confE...1K}.

The images show a clear difference between the cases with and without X-ray heating even at these low resolutions, suggesting that SKA should be able to distinguish between these two scenarios. As the variations over the field of view (FoV) reach values of 50~mK we can also conclude that SKA1-Low will be able to image these structures for deep integrations of around 1000 hours. Previous expectations were that SKA1-Low would only be able to make statistical detections of the 21-cm signal from the Cosmic Dawn. Our results indicate that, at least from the perspective of signal to noise, imaging should be possible.

Once again, the signal from partially heated IGM peaks around $z\sim15$, when there are large regions -- tens of Mpc across -- in either emission or absorption. This peak is followed by gradual thorough heating of the IGM above $T_{\mathrm{CMB}}$, bringing the signal into emission and, thus, decreasing the overall fluctuations. In the stellar-only case, the maps remain fully in absorption at these resolutions, since the ionized regions are much smaller than the beam extent and are smoothed away. Nonetheless, considerable fluctuations remain, with high overall amplitude due to the very cold IGM in that case. Over time larger regions are substantially affected by further structure formation occurs the contrast in the images becomes greater.

\subsection{21-CM ONE-POINT STATISTICS} 
\label{sec:21-cm}

The 21-cm fluctuations are typically non-Gaussian in nature and are not fully described by the power spectra alone, and imaging might not be possible in all regimes for sensitivity reasons. Therefore, the one-point statistical properties of the 21-cm signal are also of great interest, since they quantify other aspects of the 21-cm signal and enable comparisons with past works and future observations. The rms is defined as:

\begin{equation}
\mathrm{rms}(y) \equiv \sigma=\sqrt{\frac{\sum_{i=0}^{N}(y_i - \overline{y})^2}{N}},
\label{eq:rms}
\end{equation}
and we use the dimensionless definitions for the skewness and kurtosis, as follows:

\begin{equation}
\mathrm{Skewness}(y) = \frac{1}{N} \frac{\sum_{i=0}^{N}(y_i - \overline{y})^3}{\sigma^{3}},
\label{eq:skew}
\end{equation}
and
\begin{equation}
\mathrm{Kurtosis}(y) = \frac{1}{N} \frac{\sum_{i=0}^{N}(y_i - \overline{y})^4}{\sigma^4}.
\label{eq:kur}
\end{equation}

\begin{figure*} 
\centering
\makebox[\textwidth][c]{\includegraphics[width=1.\textwidth]{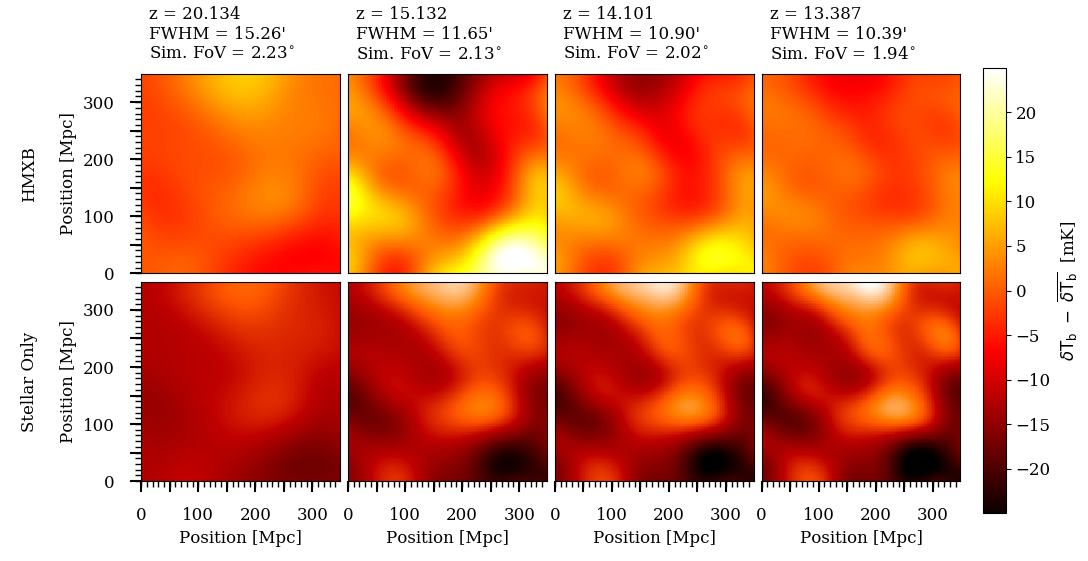}
}
\caption{Mean-subtracted $\delta T_{\rm b}$ maps smoothed with a Gaussian beam with the  FWHM corresponding to a 1.2~km maximum baseline at the relevant frequency, as labelled. The images are bandwidth-smoothed with a top hat function (width equal to the distance corresponding to the beam width). The X-ray simulation runs along the top row and the stellar-only case is below, with snapshots of the same redshifts being vertically aligned.}
\label{fig:dbtmaps}
\end{figure*}

Here, $y$ is the quantity of interest (in this case $\delta T_{\rm b}$), $N$ is the number of data points, $\overline{y}$ is the mean value of $y$, and $\sigma^2$ is the variance of $y$. These quantities are calculated from coeval simulation cubes, smoothed with a Gaussian beam corresponding to a 1.2~km maximum baseline at the relevant frequency and a bandwidth corresponding to the same spatial extent as the full-width half-maximum (FWHM) of the beam. This smoothing is the same as was used for the images in Fig.~\ref{fig:dbtmaps}.

\begin{figure*} 
\centering
\makebox[\textwidth][c]{\includegraphics[width=1.0\textwidth]{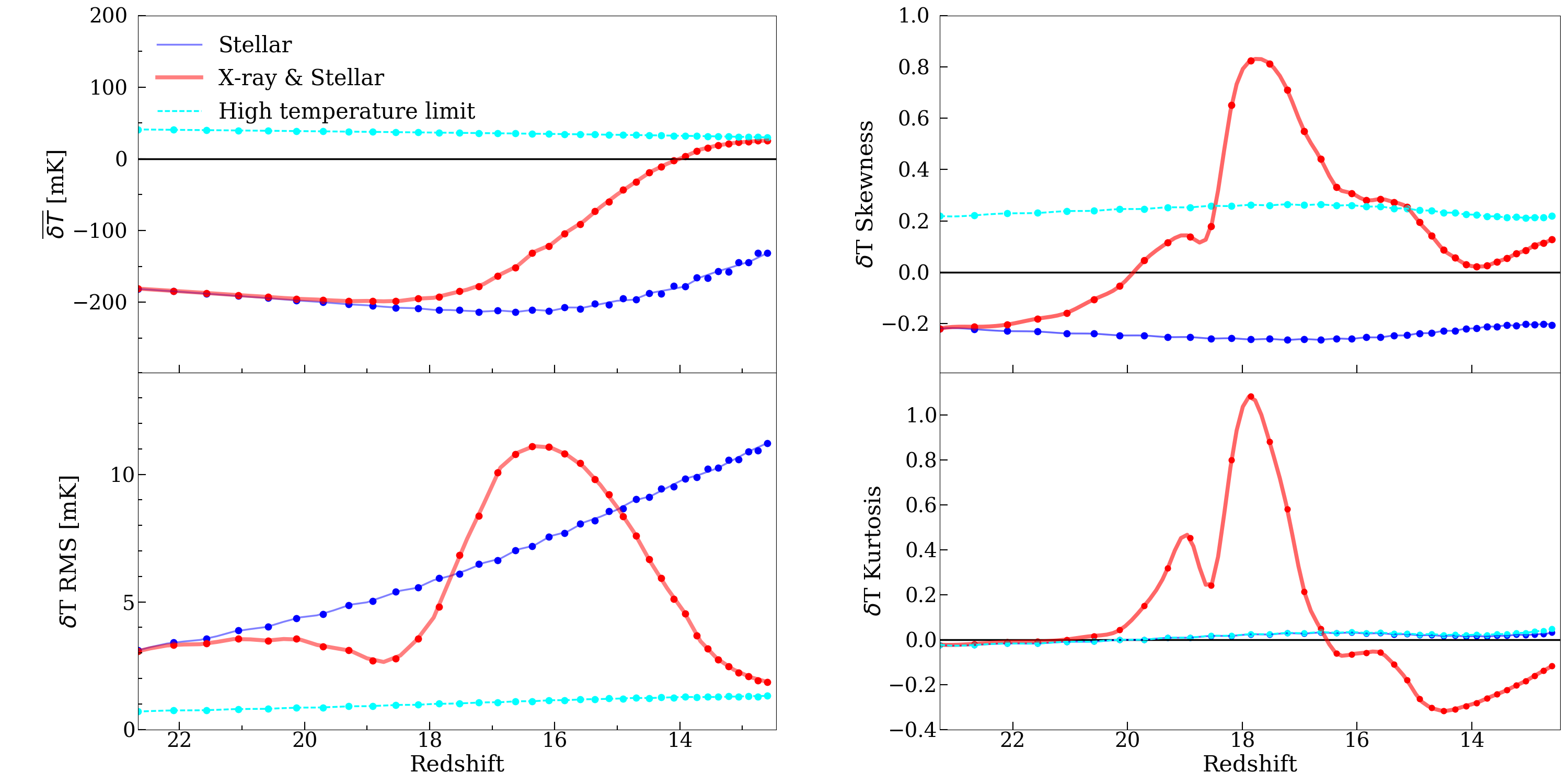}}
\caption{Statistics from the 21-cm signal from both our simulations as well as the high-$T$$_\mathrm{K}$ limit. The top left panel shows the mean value of $\delta T_{\rm b}$, the bottom left panel the rms, the top-right panel the skewness and the bottom right the kurtosis. The points are the results calculated from smoothed coeval boxes from our simulations and the fitted line is a cubic spline of these data.}
\label{fig:statistics}
\end{figure*}

In the top-left panel of Fig. \ref{fig:statistics}, we show the global value of the $\delta T_{\rm b}$  calculated as the mean signal from our simulations, $\overline{ \delta T_\mathrm{b}}$. The high-$T_\mathrm{K}$ limit corresponds to the dashed (cyan) lines. In both simulation models, the global value starts negative, due to the initially cold IGM, and drops further as the universe expands and cools adiabatically. In the stellar-only case (thin, blue line), $\overline{ \delta T_\mathrm{b}}$ rises slowly thereafter, starting at $z\sim16.5$ as the highest density peaks become ionized. The highest value remains negative, since the neutral IGM never gets heated in this scenario. $\overline{ \delta T_\mathrm{b}}$ is significantly higher in the HMXB case (thick, red line), starting to rise from around $z = 20$ due to the heating of the IGM. The global value becomes positive just before $z=14$, and by the end of the simulation it approaches the high-$T$$_\mathrm{K}$ limit. The evolution of the global 21-cm signal is similar to that of the analytical and semi-numerical models in the literature \citep[e.g.][]{Pritchard2007,Mesinger2013}, apart from the timing of this transition, which depends on the specific assumptions made about the ionizing and X-ray sources. One difference from these models is that we do not model the early \Lya\  background, but assume efficient WF coupling at all times. When this assumption is not made, weaker WF coupling early on produces a shallower absorption signal (typically $\overline{ \delta T_\mathrm{b}}_{\rm min}>180\,\rm mK$ instead of $\overline{ \delta T_\mathrm{b}}_{\rm min}\sim200\,\rm mK$ as in our case). While the incomplete \Lya\  coupling could be an important effect at the earliest times ($z>20$), we focus on the subsequent X-ray heating epoch, where this effect should have minimal impact.

The lower left-hand panel of Fig. \ref{fig:statistics} shows the rms or standard deviation of the $\delta T_\mathrm{b}$ as a function of redshift, calculated according to Eq.~\ref{eq:rms} and for the resolution specified below Eq.~\ref{eq:kur}. Before the X-ray heating is able to significantly impact the cold IGM, our two scenarios have a similar rms evolution, which is dominated by the density fluctuations and the adiabatic cooling of the IGM. Later on, the rms drops slightly in the HMXB case due to the local X-ray heating from the first sources, which brings the local IGM temperature closer to that of the CMB. As the characteristic scale of the X-ray heating fluctuations increases (cf. the power spectra evolution), the rms of the X-ray case starts rising again and peaks at $\sim 11.5$~mK around $z\sim16.5$. Thereafter, the rms fluctuations gradually decrease due to the $\overline{\delta T_\mathrm{b}}$ rising towards positive values and the 21-cm absorption turning into emission. The rms in the stellar-only case continuously rises, as the density fluctuations increase. Although local ionization introduces small-scale $\delta T_\mathrm{b}$ fluctuations, they are smoothed out by the beam- and bandwidth-averaging.
 
At later times in the HMXB case, the rms asymptotically approaches the high-$T$$_\mathrm{K}$ limit, but does not quite reach it by the end of the simulation at $z=12.7$. In the stellar-only case, the rms of $\delta T_\mathrm{b}$ is driven purely by the density (and later on to a small extent ionization) fluctuations. The rms, therefore, continues to rise as structure formation continues, ionization begins, and the IGM further cools. Note that in reality the temperature of the neutral IGM is likely to be impacted by shock heating, and we leave this to future work. The features of the rms evolution and their timing are dependent on the resolution available and, hence, on the details of the radio interferometer.

The higher order statistics of $\delta T_{\rm b}$ are also affected by the inclusion of X-rays. The skewness of $\delta T_{\rm b}$ is shown in the top-right panel of Fig.~\ref{fig:statistics}. In both cases, the skewness starts close to zero, tracking the initial, Gaussian density fields. The skewness then gradually increases in the HMXB case as hot regions surrounding sources positively skew the data. The skewness from the HMXB case then increases rapidly, as large regions of the IGM are heated. The skewness then peaks around $z \sim 18$ with a value of 0.85. After this peak, the skewness decreases as the heating becomes more homogeneous and reaches zero before increasing again and approaching the high-$T_\mathrm{K}$ limit (shown as dashed, cyan line). The skewness from the stellar-only case remains negative throughout the simulation and only begins to rise toward the end of the simulation when the ionized fraction becomes non-negligible. Once again, the high-$T_\mathrm{K}$ limit is not valid for the skewness at the early times considered in this work, except for the very last stages of the HMXB case. The skewness from the high-$T_\mathrm{K}$ limit and the stellar-only case are mirror images of each other due to the fact that they are both dominated by density fluctuations, as $T_{\rm s} \gg T_{\rm CMB}$ in the high-$T_\mathrm{K}$ limit and the stellar-only case is dominated by the adiabatic temperature of the universe. The high-$T_\mathrm{K}$ limit is forced to be in emission; whereas the stellar-only case is in absorption. Our skewness results are in reasonable agreement with the ones for low X-ray efficiency case `$\log\zeta_{\rm X}=55$' in \citet{2015MNRAS.454.1416W} (cf. their Fig.~13, but note that they show the dimensional skewness, which is the same as the dimensionless skewness multiplied by the corresponding rms).

The kurtosis of $\delta T_{\rm b}$ from our two simulations is displayed in the bottom right panel of Fig.~\ref{fig:statistics}. Initially, the kurtosis of both models is close to zero as the Gaussian density fluctuations dominate at these early stages. As long-range heating develops, the kurtosis of the HMXB case increases and peaks at $z \sim 18$ with a value of 1.1. Later, the kurtosis decreases again to negative values, before increasing once more and approaching temperature saturation (shown as dashed, cyan line). This statistic follows roughly the same pattern as the skewness, but with somewhat different functional shape and timing. The stellar-only case kurtosis closely tracks that of the high-$T_\mathrm{K}$ limit throughout the simulation, deviating only slightly at the end due to the small amount of ionization present. As there is no heating in this case, the signal is mostly Gaussian with small non-Gaussianity only arising from the density fluctuations.

\section{CONCLUSIONS}
\label{sec:conclusions}
We present the first large-volume, fully numerical structure formation and radiative transfer simulations of the IGM heating during the Cosmic Dawn by the first stellar and X-ray sources. We simulate the multi-frequency transfer of both ionizing and X-ray photons and solve self-consistently for the temperature state of the IGM. While the exact nature and properties of the first X-ray sources are still quite uncertain, our results demonstrate that, under a reasonable set of assumptions, these sources produce significant early and inhomogeneous heating of the neutral IGM and, thus, impact considerably the redshifted 21-cm signals. We focus on these expected 21-cm signals from this epoch and its statistics throughout this paper.

In this work, we consider relatively soft-spectrum X-ray sources, which trace the star formation at high redshift. At these high redshifts, these sources are still fairly rare and for reasonable assumed efficiencies, the addition of X-rays does not affect significantly the evolution of the mean fractions of H~II and He~II. The fraction of He~III, however, is boosted by almost an order of magnitude compared to the stellar-only case, although remaining quite low overall.

The high energies and long mean free paths of the hard X-ray radiation make it the dominant driver of the heating of the neutral IGM. Pop.~II stars, even massive ones, do not produce a significant amount of such hard radiation. Therefore, both the morphology and the overall amount of heating change dramatically when X-ray sources are present. The mean and the median temperature both increase considerably compared to the stellar-only case, with the mean eventually reaching $\sim10^3$~K by $z\sim13$ (the median, which only reaches $\sim200$~K, better reflects the neutral IGM state as it is less sensitive to the very high temperatures in the ionized regions). The X-ray heating is long-range and, therefore, widely distributed throughout the IGM. This heating is also highly inhomogeneous, as evidenced by the temperature PDFs, maps, and evolution seen in the lightcone visualisations. The neutral regions are heated by the X-ray sources and go fully into 21-cm emission with respect to the CMB before $z=13$, while with stellar-only sources the IGM remains in absorption throughout the Cosmic Dawn. The presence of X-rays, therefore, results in an early, but extended ($\Delta z\sim 7$) transition into emission.

The 21-cm fluctuations initially ($z>20$) track the density fluctuations due to the still insignificant heating and ionization fluctuations. However, the temperature fluctuations due to X-ray heating quickly boost the large-scale 21-cm fluctuations to much higher values. At a resolution of $\sim 10-12$ arcmin for redshifts 15 -- 17, the fluctuations are large enough to be a factor of several above the expected noise level of SKA1-Low, which implies the possibility of observing not only power spectra, but also coarse images of the 21-cm signal from the Cosmic Dawn. For the same resolution, the $\delta T_{\rm b}$ rms in the presence of X-rays peaks at $\sim11.5$~mK around $z\sim16.5$. As the X-rays heat the neutral IGM, a broad peak develops at $k\sim 0.1$~Mpc$^{-1}$, corresponding to spatial scale of about 43 Mpc, at $z \sim14-15$. As the IGM heats up and the absorption gradually turns into emission, the 21-cm fluctuations for the HMXB case decrease and asymptote to the high-$T_\mathrm{K}$ limit, which is not fully reached by the end of our simulation ($z\sim12.7$), even though by that time the mean IGM is heated well above the CMB temperature. In contrast, the stellar-only case fluctuations are still increasing steeply by $z\sim12$, as they are driven by cold IGM.

In the HMXB case, the distribution of the $\delta T_{\rm b}$ fluctuations shows a clear non-Gaussian signature, with both the skewness and kurtosis peaking when the fluctuations start rising. By the end of the simulation, the skewness and kurtosis approach, but do not reach, the high-$T_\mathrm{K}$ limit. For soft radiation sources, the non-Gaussianity is driven by density fluctuations only, producing a smooth evolution.

The often-used high spin temperature limit, $T_\mathrm{S} \gg T_{\mathrm{CMB}}$, is not valid throughout the X-ray heating epoch as long as any IGM patches remain cold. When X-rays are present, even after the IGM temperature rises above the CMB everywhere (and thus the 21-cm signal transits into emission), significant temperature fluctuations remain and contribute to the 21-cm signal. The neutral regions do not asymptote to the high-temperature limit until quite late in our model, at $z\sim12$. This asymptotic behavior can readily been seen in the power spectra and statistics of the 21-cm signal. Soft, stellar-only radiation has short mean free paths and, therefore, never penetrates into the neutral regions, leaving a cold IGM.

Previous work in this area has largely been limited to approximate semi-analytical and semi-numerical modelling \citep[e.g.][]{Pritchard2007,Mesinger2013,2014MNRAS.445..213F,2015MNRAS.451..467S,
2015MNRAS.454.1416W,2016arXiv160202873K}. By their nature, such approaches do not apply detailed, multi-frequency RT, but rely on counting the photons produced in a certain region of space and comparing this to the number of atoms (with some correction for the recombinations occurring). The difference between the two determines the ionization state of that region.
The X-ray heating is done by solving the energy equation using integrated, average optical depths and photon fluxes, and often additional approximations are employed as well \citep[e.g.][]{2015MNRAS.454.1416W}. These methods typically do not take into account nonlinear physics, spatially varying gas clumping or absorbers, or Jeans mass filtering of low mass sources. These differences make comparisons with the previous results in detail difficult due to the very different modelling employed and would require further study. Nonetheless, we find some commonalities and some disparities with our results, summarised below. 

Our thermal history is similar to that of the relevant cases in \citet{Pritchard2007} (their Case A) and \citet{2015MNRAS.454.1416W} (their case `$\log\zeta_{\rm X}=55$'). We find a quite extended transition between 21-cm absorption and emission, from the formation of the first ionizing and X-ray sources at $z\sim21$ all the way to $z\sim13$. This transition is somewhat more protracted than the one in the most similar scenarios ($f_{\rm x}=1$ and $5$) considered in \citet{Mesinger2013}, likely due to the higher star formation efficiencies assumed in that work. 

We find a clear X-ray heating-driven peak in the 21-cm power spectra at $k=0.1-0.2\,{\rm Mpc}^{-1}$, similar to the soft X-ray spectra peak found in\citet{2014MNRAS.443..678P} and at similar redshift (z$\sim 19 \ 15 - 16$; though this depends on the uncertain source efficiencies). Results from their peak power,  at $\Delta_{\rm 21cm}\sim14\,$mK, are in rough agreement with our results. The general evolution of the power spectra found in \cite{Pritchard2007} appears similar, with the fluctuations at $k=0.1\,\rm Mpc^{-1}$ also peaking at $z\sim 15-16$ (although that only occurs at $z\sim 12-13$ for the scenario with less X-rays, again suggesting a strong dependence on the source model). The power spectra found are in reasonable agreement our results, with peak values of $\Delta_{\rm 21cm}\sim19\,$mK or $\Delta_{\rm 21cm}\sim11.5\,$mK depending on the source model used by them, compared to $\Delta_{\rm 21cm}\sim14\,$mK for our HMXB case.

The 21-cm skewness from the X-ray heating epoch is rarely calculated, but \citet{2015MNRAS.454.1416W} recently found a very similar evolution to ours (though shifted to somewhat higher redshifts), with a positive peak roughly coinciding with the initial rise of the 21-cm fluctuations due to the temperature patchiness. Their corresponding 21-cm $\delta T_{\rm b}$ PDF distributions during the X-ray heating epoch significantly differ from ours, however. At the epoch when $T_{\rm S}$ reaches a minimum, the semi-numerical model predicts a long tail of positive $\delta T_{\rm b}$, which does not exist in the full simulations. Around the 
$T_{\rm S}\sim T_{\rm CMB}$ epoch, our distribution is quite Gaussian; while \citet{2015MNRAS.454.1416W} find an asymmetric one (though, curiously, with one with close to zero skewness, indicating that skewness alone provides a very incomplete description). Finally, in the $T_{\rm S}\gg T_{\rm CMB}$ epoch, the two results both yield gaussian PDFs, but the simulated one is much narrower.

Our models confirm that, for reasonable assumptions about the presence of X-ray sources, there is a period of substantial of fluctuations in the 21-cm signal caused by the patchiness of this heating and that this period precedes the one in which fluctuations are mostly caused by patchiness in the ionization. However, since the nature and properties of X-ray sources remain unconstrained by observations, other scenarios in which the heating occurs later, are also allowed. The currently ongoing observational campaigns of both LOFAR and MWA should be able to put constraints on the presence of spin temperature fluctuations for the range $z < 11$, which would then have clear implications for the required efficiency of X-ray heating at those and earlier redshifts. In the future, we will use simulations of the kind presented here to explore other possible scenarios, for example heating caused by rare, bright sources, as well as the impact of spin temperature fluctuations on all aspects of the 21-cm signal, such as redshift space distortions.

\section{Acknowledgements}
This work was supported by the Science and Technology Facilities Council [grant number ST/I000976/1] and the Southeast Physics Network (SEPNet). GM is supported in part by Swedish Research Council grant 2012-4144. This research was supported in part by the Munich Institute for Astro- and Particle Physics (MIAPP) of the DFG cluster of excellence ``Origin and Structure of the Universe". We acknowledge that the results in this paper have been achieved using the PRACE Research Infrastructure resources Curie based at the Très Grand Centre de Calcul (TGCC) operated by CEA near Paris, France and Marenostrum based in the Barcelona Supercomputing Center, Spain. Time on these resources was awarded by PRACE under PRACE4LOFAR grants 2012061089 and 2014102339 as well as under the Multi-scale Reionization grants 2014102281 and 2015122822. Some of the numerical computations were done on the Apollo cluster at The University of Sussex.
\newpage 
\bibliography{paper}

\begin{thebibliography}{52}
\expandafter\ifx\csname natexlab\endcsname\relax\def\natexlab#1{#1}\fi

\bibitem[{{Ahn} {et~al.}(2015{\natexlab{a}}){Ahn}, {Iliev}, {Shapiro}, \&
  {Srisawat}}]{Ahn15a}
{Ahn} K., {Iliev} I.~T., {Shapiro} P.~R., {Srisawat} C., 2015{\natexlab{a}},
  \mnras, 450, 1486

\bibitem[{{Ahn} {et~al.}(2015{\natexlab{b}}){Ahn}, {Xu}, {Norman}, {Alvarez},
  \& {Wise}}]{Ahn2015}
{Ahn} K., {Xu} H., {Norman} M.~L., {Alvarez} M.~A., {Wise} J.~H.,
  2015{\natexlab{b}}, \apj, 802, 8

\bibitem[{{Baek} {et~al.}(2010){Baek}, {Semelin}, {Di Matteo}, {Revaz}, \&
  {Combes}}]{Baek2010}
{Baek} S., {Semelin} B., {Di Matteo} P., {Revaz} Y., {Combes} F., 2010, \aa,
  523

\bibitem[{{Bolton} {et~al.}(2012){Bolton}, {Becker}, {Raskutti}, {Wyithe},
  {Haehnelt}, \& {Sargent}}]{Bolton2012}
{Bolton} J.~S., {Becker} G.~D., {Raskutti} S., {Wyithe} J.~S.~B., {Haehnelt}
  M.~G., {Sargent} W.~L.~W., 2012, \mnras, 419, 2880

\bibitem[{{Ciardi} \& {Madau}(2003)}]{2003ApJ...596....1C}
{Ciardi} B., {Madau} P., 2003, \apj, 596, 1

\bibitem[{{Crocce} {et~al.}(2006){Crocce}, {Pueblas}, \&
  {Scoccimarro}}]{Crocce2006}
{Crocce} M., {Pueblas} S., {Scoccimarro} R., 2006, \mnras, 373, 369

\bibitem[{{Dixon} {et~al.}(2016){Dixon}, {Iliev}, {Mellema}, {Ahn}, \&
  {Shapiro}}]{2016MNRAS.456.3011D}
{Dixon} K.~L., {Iliev} I.~T., {Mellema} G., {Ahn} K., {Shapiro} P.~R., 2016,
  \mnras, 456, 3011

\bibitem[{{Fan} {et~al.}(2006){Fan}, {Strauss}, {Becker}, {White}, {Gunn},
  {Knapp}, {Richards}, {Schneider}, {Brinkmann}, \& {Fukugita}}]{Fan2006}
{Fan} X., {Strauss} M.~A., {Becker} R.~H., {White} R.~L., {Gunn} J.~E., {Knapp}
  G.~R., {Richards} G.~T., {Schneider} D.~P., {Brinkmann} J., {Fukugita} M.,
  2006, \aj, 132, 117

\bibitem[{{Fialkov} \& {Barkana}(2014)}]{2014MNRAS.445..213F}
{Fialkov} A., {Barkana} R., 2014, \mnras, 445, 213

\bibitem[{{Fialkov} {et~al.}(2014){Fialkov}, {Barkana}, \&
  {Visbal}}]{Fialkov2014}
{Fialkov} A., {Barkana} R., {Visbal} E., 2014, \nat, 506, 197

\bibitem[{{Field}(1958)}]{Field1958}
{Field} G.~B., 1958, Proceedings of the IRE, 46, 240

\bibitem[{{Friedrich} {et~al.}(2012){Friedrich}, {Mellema}, {Iliev}, \&
  {Shapiro}}]{Friedrich2012}
{Friedrich} M.~M., {Mellema} G., {Iliev} I.~T., {Shapiro} P.~R., 2012, \mnras,
  421, 2232

\bibitem[{Furlanetto {et~al.}(2004)Furlanetto, Hernquist, \&
  Zaldarriaga}]{Furlanetto2004}
Furlanetto S.~R., Hernquist L., Zaldarriaga M., 2004, \mnras, 354, 695

\bibitem[{{Glover} \& {Brand}(2003)}]{Glover2003}
{Glover} S.~C.~O., {Brand} P.~W.~J.~L., 2003, \mnras, 340, 210

\bibitem[{{Greiner} {et~al.}(2009){Greiner}, {Kr{\"u}hler}, {Fynbo}, {Rossi},
  {Schwarz}, {Klose}, {Savaglio}, {Tanvir}, {McBreen}, {Totani}, {Zhang}, {Wu},
  {Watson}, {Barthelmy}, {Beardmore}, {Ferrero}, {Gehrels}, {Kann}, {Kawai},
  {Yolda{\c s}}, {M{\'e}sz{\'a}ros}, {Milvang-Jensen}, {Oates}, {Pierini},
  {Schady}, {Toma}, {Vreeswijk}, {Yolda{\c s}}, {Zhang}, {Afonso}, {Aoki},
  {Burrows}, {Clemens}, {Filgas}, {Haiman}, {Hartmann}, {Hasinger}, {Hjorth},
  {Jehin}, {Levan}, {Liang}, {Malesani}, {Pyo}, {Schulze}, {Szokoly}, {Terada},
  \& {Wiersema}}]{Greiner2009}
{Greiner} J., {Kr{\"u}hler} T., {Fynbo} J.~P.~U., {Rossi} A., {Schwarz} R.,
  {Klose} S., {Savaglio} S., {Tanvir} N.~R., {McBreen} S., {Totani} T., {Zhang}
  B.~B., {Wu} X.~F., {Watson} D., {Barthelmy} S.~D., {Beardmore} A.~P.,
  {Ferrero} P., {Gehrels} N., {Kann} D.~A., {Kawai} N., {Yolda{\c s}} A.~K.,
  {M{\'e}sz{\'a}ros} P., {Milvang-Jensen} B., {Oates} S.~R., {Pierini} D.,
  {Schady} P., {Toma} K., {Vreeswijk} P.~M., {Yolda{\c s}} A., {Zhang} B.,
  {Afonso} P., {Aoki} K., {Burrows} D.~N., {Clemens} C., {Filgas} R., {Haiman}
  Z., {Hartmann} D.~H., {Hasinger} G., {Hjorth} J., {Jehin} E., {Levan} A.~J.,
  {Liang} E.~W., {Malesani} D., {Pyo} T.-S., {Schulze} S., {Szokoly} G.,
  {Terada} K., {Wiersema} K., 2009, \apj, 693, 1610

\bibitem[{{Haiman} \& {Holder}(2003)}]{Haim03a}
{Haiman} Z., {Holder} G.~P., 2003, \apj, 595, 1

\bibitem[{{Harnois-D{\'e}raps} {et~al.}(2013){Harnois-D{\'e}raps}, {Pen},
  {Iliev}, {Merz}, {Emberson}, \& {Desjacques}}]{Harnois2013}
{Harnois-D{\'e}raps} J., {Pen} U.-L., {Iliev} I.~T., {Merz} H., {Emberson}
  J.~D., {Desjacques} V., 2013, \mnras, 436, 540

\bibitem[{{Hickox} \& {Markevitch}(2007)}]{Hickox2007}
{Hickox} R.~C., {Markevitch} M., 2007, \apjl, 661, L117

\bibitem[{{Iliev} {et~al.}(2006){Iliev}, {Ciardi}, {Alvarez}, {Maselli},
  {Ferrara}, {Gnedin}, {Mellema}, {Nakamoto}, {Norman}, {Razoumov},
  {Rijkhorst}, {Ritzerveld}, {Shapiro}, {Susa}, {Umemura}, \&
  {Whalen}}]{Iliev2006b}
{Iliev} I.~T., {Ciardi} B., {Alvarez} M.~A., {Maselli} A., {Ferrara} A.,
  {Gnedin} N.~Y., {Mellema} G., {Nakamoto} T., {Norman} M.~L., {Razoumov}
  A.~O., {Rijkhorst} E.-J., {Ritzerveld} J., {Shapiro} P.~R., {Susa} H.,
  {Umemura} M., {Whalen} D.~J., 2006, \mnras, 371, 1057

\bibitem[{{Iliev} {et~al.}(2014){Iliev}, {Mellema}, {Ahn}, {Shapiro}, {Mao}, \&
  {Pen}}]{Iliev2013}
{Iliev} I.~T., {Mellema} G., {Ahn} K., {Shapiro} P.~R., {Mao} Y., {Pen} U.-L.,
  2014, \mnras, 439, 725

\bibitem[{{Iliev} {et~al.}(2007){Iliev}, {Mellema}, {Shapiro}, \&
  {Pen}}]{2007MNRAS.376..534I}
{Iliev} I.~T., {Mellema} G., {Shapiro} P.~R., {Pen} U.-L., 2007, \mnras, 376,
  534

\bibitem[{{Iliev} {et~al.}(2012){Iliev}, {Mellema}, {Shapiro}, {Pen}, {Mao},
  {Koda}, \& {Ahn}}]{Ilie12a}
{Iliev} I.~T., {Mellema} G., {Shapiro} P.~R., {Pen} U.-L., {Mao} Y., {Koda} J.,
  {Ahn} K., 2012, \mnras, 423, 2222

\bibitem[{{Iliev} {et~al.}(2010){Iliev}, {Whalen}, {Mellema}, {Ahn}, {Baek},
  {Gnedin}, {Kravtsov}, {Norman}, {Raicevic}, {Reynolds}, {Sato}, {Shapiro},
  {Semelin}, {Smidt}, {Susa}, {Theuns}, \& {Umemura}}]{Iliev2010}
{Iliev} I.~T., {Whalen} D., {Mellema} G., {Ahn} K., {Baek} S., {Gnedin} N.~Y.,
  {Kravtsov} A.~V., {Norman} M., {Raicevic} M., {Reynolds} D.~R., {Sato} D.,
  {Shapiro} P.~R., {Semelin} B., {Smidt} J., {Susa} H., {Theuns} T., {Umemura}
  M., 2010, \mnras, 400, 1283

\bibitem[{{Khaire} {et~al.}(2016){Khaire}, {Srianand}, {Choudhury}, \&
  {Gaikwad}}]{Khaire2016}
{Khaire} V., {Srianand} R., {Choudhury} T.~R., {Gaikwad} P., 2016, \mnras, 457,
  4051

\bibitem[{{Knevitt} {et~al.}(2014){Knevitt}, {Wynn}, {Power}, \&
  {Bolton}}]{Knevitt2014}
{Knevitt} G., {Wynn} G.~A., {Power} C., {Bolton} J.~S., 2014, \mnras, 445, 2034

\bibitem[{{Komatsu} {et~al.}(2011){Komatsu}, {Smith}, {Dunkley}, {Bennett},
  {Gold}, {Hinshaw}, {Jarosik}, {Larson}, {Nolta}, {Page}, {Spergel},
  {Halpern}, {Hill}, {Kogut}, {Limon}, {Meyer}, {Odegard}, {Tucker}, {Weiland},
  {Wollack}, \& {Wright}}]{Komatsu2011}
{Komatsu} E., {Smith} K.~M., {Dunkley} J., {Bennett} C.~L., {Gold} B.,
  {Hinshaw} G., {Jarosik} N., {Larson} D., {Nolta} M.~R., {Page} L., {Spergel}
  D.~N., {Halpern} M., {Hill} R.~S., {Kogut} A., {Limon} M., {Meyer} S.~S.,
  {Odegard} N., {Tucker} G.~S., {Weiland} J.~L., {Wollack} E., {Wright} E.~L.,
  2011, \apjs, 192, 18

\bibitem[{{Koopmans} {et~al.}(2015){Koopmans}, {Pritchard}, {Mellema},
  {Aguirre}, {Ahn}, {Barkana}, {van Bemmel}, {Bernardi}, {Bonaldi}, {Briggs},
  {de Bruyn}, {Chang}, {Chapman}, {Chen}, {Ciardi}, {Dayal}, {Ferrara},
  {Fialkov}, {Fiore}, {Ichiki}, {Illiev}, {Inoue}, {Jelic}, {Jones}, {Lazio},
  {Maio}, {Majumdar}, {Mack}, {Mesinger}, {Morales}, {Parsons}, {Pen},
  {Santos}, {Schneider}, {Semelin}, {de Souza}, {Subrahmanyan}, {Takeuchi},
  {Vedantham}, {Wagg}, {Webster}, {Wyithe}, {Datta}, \&
  {Trott}}]{2015aska.confE...1K}
{Koopmans} L., {Pritchard} J., {Mellema} G., {Aguirre} J., {Ahn} K., {Barkana}
  R., {van Bemmel} I., {Bernardi} G., {Bonaldi} A., {Briggs} F., {de Bruyn}
  A.~G., {Chang} T.~C., {Chapman} E., {Chen} X., {Ciardi} B., {Dayal} P.,
  {Ferrara} A., {Fialkov} A., {Fiore} F., {Ichiki} K., {Illiev} I.~T., {Inoue}
  S., {Jelic} V., {Jones} M., {Lazio} J., {Maio} U., {Majumdar} S., {Mack}
  K.~J., {Mesinger} A., {Morales} M.~F., {Parsons} A., {Pen} U.~L., {Santos}
  M., {Schneider} R., {Semelin} B., {de Souza} R.~S., {Subrahmanyan} R.,
  {Takeuchi} T., {Vedantham} H., {Wagg} J., {Webster} R., {Wyithe} S., {Datta}
  K.~K., {Trott} C., 2015, Advancing Astrophysics with the Square Kilometre
  Array (AASKA14), 1

\bibitem[{{Kubota} {et~al.}(2016){Kubota}, {Yoshiura}, {Shimabukuro}, \&
  {Takahashi}}]{2016arXiv160202873K}
{Kubota} K., {Yoshiura} S., {Shimabukuro} H., {Takahashi} K., 2016, ArXiv
  e-prints 1602.02873

\bibitem[{{Lee} {et~al.}(2016){Lee}, {Mellema}, \& {Lundqvist}}]{Lee2016}
{Lee} K.-Y., {Mellema} G., {Lundqvist} P., 2016, \mnras, 455, 4406

\bibitem[{{Lewis} {et~al.}(2000){Lewis}, {Challinor}, \& {Lasenby}}]{Lewis2000}
{Lewis} A., {Challinor} A., {Lasenby} A., 2000, \apj, 538, 473

\bibitem[{{Lutovinov} {et~al.}(2005){Lutovinov}, {Revnivtsev}, {Gilfanov},
  {Shtykovskiy}, {Molkov}, \& {Sunyaev}}]{Lutivnov2005}
{Lutovinov} A., {Revnivtsev} M., {Gilfanov} M., {Shtykovskiy} P., {Molkov} S.,
  {Sunyaev} R., 2005, \aap, 444, 821

\bibitem[{{McGreer} {et~al.}(2015){McGreer}, {Mesinger}, \&
  {D'Odorico}}]{McGreer2015}
{McGreer} I.~D., {Mesinger} A., {D'Odorico} V., 2015, \mnras, 447, 499

\bibitem[{{Mellema} {et~al.}(2006){Mellema}, {Iliev}, {Alvarez}, \&
  {Shapiro}}]{Mellema2006}
{Mellema} G., {Iliev} I.~T., {Alvarez} M.~A., {Shapiro} P.~R., 2006, Elsevier
  Science, 11, 374

\bibitem[{{Mesinger} {et~al.}(2013){Mesinger}, {Ferrara}, \&
  {Spiegel}}]{Mesinger2013}
{Mesinger} A., {Ferrara} A., {Spiegel} D.~S., 2013, \mnras, 431, 621

\bibitem[{{Mineo} {et~al.}(2012){Mineo}, {Gilfanov}, \& {Sunyaev}}]{Mineo2012}
{Mineo} S., {Gilfanov} M., {Sunyaev} R., 2012, \mnras, 419, 2095

\bibitem[{{Pacucci} {et~al.}(2014){Pacucci}, {Mesinger}, {Mineo}, \&
  {Ferrara}}]{2014MNRAS.443..678P}
{Pacucci} F., {Mesinger} A., {Mineo} S., {Ferrara} A., 2014, \mnras, 443, 678

\bibitem[{{Pentericci} {et~al.}(2014){Pentericci}, {Vanzella}, {Fontana},
  {Castellano}, {Treu}, {Mesinger}, {Dijkstra}, {Grazian}, {Brada{\v c}},
  {Conselice}, {Cristiani}, {Dunlop}, {Galametz}, {Giavalisco}, {Giallongo},
  {Koekemoer}, {McLure}, {Maiolino}, {Paris}, \& {Santini}}]{Pentericci2014}
{Pentericci} L., {Vanzella} E., {Fontana} A., {Castellano} M., {Treu} T.,
  {Mesinger} A., {Dijkstra} M., {Grazian} A., {Brada{\v c}} M., {Conselice} C.,
  {Cristiani} S., {Dunlop} J., {Galametz} A., {Giavalisco} M., {Giallongo} E.,
  {Koekemoer} A., {McLure} R., {Maiolino} R., {Paris} D., {Santini} P., 2014,
  \apj, 793, 113

\bibitem[{{Planck Collaboration} {et~al.}(2016){Planck Collaboration}, {Adam},
  {Aghanim}, {Ashdown}, {Aumont}, \& et~al}]{Planck2016}
{Planck Collaboration}, {Adam} R., {Aghanim} N., {Ashdown} M., {Aumont} J.,
  et~al B., 2016, ArXiv e-prints 1605.03507

\bibitem[{{Planck Collaboration} {et~al.}(2015){Planck Collaboration}, {Ade},
  {Aghanim}, {Arnaud}, {Ashdown}, {Aumont}, {Baccigalupi}, {Banday},
  {Barreiro}, {Bartlett}, \& et~al.}]{Planck2015}
{Planck Collaboration}, {Ade} P.~A.~R., {Aghanim} N., {Arnaud} M., {Ashdown}
  M., {Aumont} J., {Baccigalupi} C., {Banday} A.~J., {Barreiro} R.~B.,
  {Bartlett} J.~G., et~al., 2015, ArXiv e-prints 1502.01589

\bibitem[{{Pritchard} \& {Furlanetto}(2007)}]{Pritchard2007}
{Pritchard} J.~R., {Furlanetto} S.~R., 2007, \mnras, 376, 1680

\bibitem[{{Pritchard} \& {Loeb}(2012)}]{2012RPPh...75h6901P}
{Pritchard} J.~R., {Loeb} A., 2012, Reports on Progress in Physics, 75, 086901

\bibitem[{{Raga} {et~al.}(1999){Raga}, {Mellema}, {Arthur}, {Binette},
  {Ferruit}, \& {Steffen}}]{1999RMxAA..35..123R}
{Raga} A.~C., {Mellema} G., {Arthur} S.~J., {Binette} L., {Ferruit} P.,
  {Steffen} W., 1999, Revista Mexicana de Astronomia y Astrofisica, 35, 123

\bibitem[{{Raskutti} {et~al.}(2012){Raskutti}, {Bolton}, {Wyithe}, \&
  {Becker}}]{Raskutti2012}
{Raskutti} S., {Bolton} J.~S., {Wyithe} J.~S.~B., {Becker} G.~D., 2012, \mnras,
  421, 1969

\bibitem[{{Santos} {et~al.}(2010){Santos}, {Ferramacho}, {Silva}, {Amblard}, \&
  {Cooray}}]{Santos2010}
{Santos} M.~G., {Ferramacho} L., {Silva} M.~B., {Amblard} A., {Cooray} A.,
  2010, \mnras, 406, 2421

\bibitem[{{Schenker} {et~al.}(2012){Schenker}, {Stark}, {Ellis}, {Robertson},
  {Dunlop}, {McLure}, {Kneib}, \& {Richard}}]{Schenker2012}
{Schenker} M.~A., {Stark} D.~P., {Ellis} R.~S., {Robertson} B.~E., {Dunlop}
  J.~S., {McLure} R.~J., {Kneib} J.-P., {Richard} J., 2012, \apj, 744, 179

\bibitem[{{Shimabukuro} {et~al.}(2015){Shimabukuro}, {Yoshiura}, {Takahashi},
  {Yokoyama}, \& {Ichiki}}]{2015MNRAS.451..467S}
{Shimabukuro} H., {Yoshiura} S., {Takahashi} K., {Yokoyama} S., {Ichiki} K.,
  2015, \mnras, 451, 467

\bibitem[{{Stark} {et~al.}(2011){Stark}, {Ellis}, \& {Ouchi}}]{Stark2011}
{Stark} D.~P., {Ellis} R.~S., {Ouchi} M., 2011, \apjl, 728, L2

\bibitem[{{Theuns} {et~al.}(2002){Theuns}, {Bernardi}, {Frieman}, {Hewett},
  {Schaye}, {Sheth}, \& {Subbarao}}]{Theuns2002}
{Theuns} T., {Bernardi} M., {Frieman} J., {Hewett} P., {Schaye} J., {Sheth}
  R.~K., {Subbarao} M., 2002, \apjl, 574, L111

\bibitem[{{Tilvi} {et~al.}(2014){Tilvi}, {Papovich}, {Finkelstein}, {Long},
  {Song}, {Dickinson}, {Ferguson}, {Koekemoer}, {Giavalisco}, \&
  {Mobasher}}]{Tilvi2014}
{Tilvi} V., {Papovich} C., {Finkelstein} S.~L., {Long} J., {Song} M.,
  {Dickinson} M., {Ferguson} H.~C., {Koekemoer} A.~M., {Giavalisco} M.,
  {Mobasher} B., 2014, \apj, 794, 5

\bibitem[{{Totani} {et~al.}(2006){Totani}, {Kawai}, {Kosugi}, {Aoki}, {Yamada},
  {Iye}, {Ohta}, \& {Hattori}}]{Totani2006}
{Totani} T., {Kawai} N., {Kosugi} G., {Aoki} K., {Yamada} T., {Iye} M., {Ohta}
  K., {Hattori} T., 2006, in IAU Joint Discussion, Vol.~7, IAU Joint Discussion

\bibitem[{{Watkinson} \& {Pritchard}(2015)}]{2015MNRAS.454.1416W}
{Watkinson} C.~A., {Pritchard} J.~R., 2015, \mnras, 454, 1416

\bibitem[{{Xu} {et~al.}(2014){Xu}, {Ahn}, {Wise}, {Norman}, \&
  {O'Shea}}]{Xu2014}
{Xu} H., {Ahn} K., {Wise} J.~H., {Norman} M.~L., {O'Shea} B.~W., 2014, \apj,
  791, 110

\end{thebibliography}
\bibliographystyle{mn} 
\end{document}